\documentclass[fleqn,usenatbib]{mnras}
\usepackage{newtxtext,newtxmath}
\usepackage{graphicx,color}
\usepackage[T1]{fontenc}
\usepackage{ae,aecompl}
\usepackage[switch]{lineno}
\usepackage{subfigure}
\usepackage[euler]{textgreek}
\usepackage{tikz}
\usepackage{array}
\usepackage{siunitx,xspace}
\usepackage{booktabs}

\newcommand{\rhb}[1]{\left\llbracket #1 \right\rrbracket }

\renewcommand{\dot}[1]{\stackrel{.}{#1}}
\renewcommand{\hat}[1]{\stackrel{\wedge}{#1}}

\renewcommand{\deg}{\ensuremath{^{\circ}}}

\newcommand{\sfl}{single-fluid\xspace}

\newcommand{\LC}{\textlambda~Cephei\xspace}

\newcommand{\Ha}{H\textalpha\xspace}
\newcommand{\ism}{_{\rm ism}}
\newcommand{\sw}{_{\rm sw}}
\newcommand{\TS}{_{\rm  TS}}
\newcommand{\rmc}{_{\rm  c}}
\newcommand{\rmA}{_{\rm  A}}
\newcommand{\rmf}{_{\rm  f}}
\newcommand{\rms}{_{\rm  s}}
\newcommand{\Msu}{\ensuremath{M_{\mathrm{s,n},1}^{2}}}

\newcommand{\M}{\ensuremath{M^{2}}}
\newcommand{\Mq}{\ensuremath{M^{4}}}
\newcommand{\Msq}{\ensuremath{M^{6}}}
\newcommand{\tant}{\ensuremath{\tan^{2}\vartheta_{1}}}

 \newcolumntype{R}{>{$}r<{$\ \ =}}
\newcolumntype{L}{>{\hspace{-0.25cm}$}l<{$}}
\newcolumntype{X}{>{\ $}l<{$}}

\title[MHD-shock structures of astrospheres]{MHD-shock structures of astrospheres: \LC-like astrospheres}

\author[K. Scherer et al.]{K. Scherer,$^{1,2}$ \thanks{kls@tp4.rub.de} L.R. Baalmann,$^1$ H. Fichtner,$^{1,2}$ J. Kleimann,$^{1}$ D.J. Bomans,$^{2,3}$\newauthor K. Weis,$^3$ S.E.S. Ferreira,$^4$ and K. Herbst$^5$\\
$^1$Institut f\"ur Theoretische Physik IV, Ruhr-Universit\"at Bochum, 44780 Bochum, Germany\\
$^2$Research Department, Plasmas with Complex Interactions, Ruhr-Universit\"at Bochum, 44780 Bochum, Germany \\
$^3$Astronomisches Institut, Ruhr-Universit\"at Bochum, 44780 Bochum, Germany, \\
$^4$Centre for Space Research,
  North-West University, 2520 Potchefstroom, South Africa,\\
$^5$Institut f\"ur Experimentelle und Angewandte Physik, Christian-Albrechts-Universit\"at zu Kiel, 24118 Kiel, Germany
 }

\date{Received: 2020 January 22 accepted: 2020 February 11, in original form 2019 November 27 }
\pubyear{2019}

\begin{document}
\label{firstpage}
\pagerange{\pageref{firstpage}--\pageref{lastpage}}
\maketitle

\begin{abstract}
The interpretation of recent observations of bow
      shocks around O-stars and the creation of corresponding models require
      a detailed understanding of the associated (magneto-)hydrodynamic
      structures. We base our study  on three-dimensional numerical  (magneto-)hydrodynamical models, which are analyzed using the dynamically relevant parameters, in particular, the (magneto)sonic Mach numbers. The analytic Rankine-Hugoniot relation for HD and MHD are compared with those obtained by the numerical model. In that context we also show that the only distance which can be approximately determined is that of the termination shock, if it is a hydrodynamical shock. For MHD shocks the stagnation point does not, in general, lie on the inflow line, which is the line parallel to the inflow vector and passing through the star. Thus an estimate via the Bernoulli equation as in the HD case is, in general, not possible. We also show that in O-star astrospheres, distinct regions exist in which the
      fast, slow, Alfv\'enic, and sonic Mach numbers become lower than
      one, implying sub-slow magnetosonic as well as sub-fast and sub-sonic
      flows. Nevertheless, the analytic MHD Rankine Hugoniot relations can be used for further studies  of turbulence and cosmic ray modulation.
   \end{abstract}
 \begin{keywords}
Stars: winds, outflows -- Magnetohydrodynamics -- Shock waves
\end{keywords}

\section{Introduction}

The interpretation of recent H\textalpha~observations of bow shocks around O-stars
\citep[][]{Meyer-etal-2016,Kobulnicky-etal-2017} and  their associated X-ray observations
\citep[e.g.][]{DeBecker-etal-2017} require a detailed analysis of the
large-scale shock structure around such stars. Simulations aiming at an
improved understanding of such observations have been performed,  e.g.\, by \citet{Decin-etal-2012}, \citet{Cox-etal-2012}, and
\citet{van-Marle-etal-2014} for
M-stars, while \citet{Arthur-2012}, 
  \citet{vanMarle-etal-2015}, \citet{Mackey-etal-2015} and
  \citet{Meyer-etal-2017} discussed the evolution of stellar wind
  bubbles.
These authors modeled astrospheres using either a hydrodynamical (HD)
or a magnetohydrodynamical (MHD) approach, either in one or two
dimensions (1D or 2D). Recently, \citet{Scherer-etal-2016a} used the
example of \LC, the brightest runaway O-star in the sky (type
O6If(n)p), to study shock structures in \sfl HD models with and
without cooling and heating.  Runaway O- and B-stars are common and
part of a sizable population in the Galaxy. A significant number of
these exhibit a bow-shock-like structure and have been discussed, e.g.\ by \citet{Huthoff-Kaper-2002},
\citet{Gvaramadze-Bomans-2008}, \citet{Gvaramadze-etal-2011}, and
\citet{Kobulnicky-etal-2010}; see also \citet{Cox-etal-2012} for the
corresponding Herschel observations in the infrared. For further references see \citet{Scherer-etal-2016a}.

Here a full 3D magnetic field is used, which causes different characteristic magnetosonic speeds because in the equatorial plane ($\vartheta=90\si{\degree}$) the Alfv\'en speed is constant, while over the poles it decays as $r^{-2}$, where $r$ is the distance from the star. In the 2D simulation by \citet{Meyer-etal-2017}, only the radial part was taken into account, which corresponds in the 3D simulation to a single line through the poles (the variation of $\varphi$ is negligible at the poles). Here we discuss an obliquely oriented ISM magnetic field, while \citet{Katushkina-etal-2018} only studied  magnetic fields either perpendicular or parallel to the flow.  In
\citet{Gvaramadze-etal-2018} the Vela-X binary is explored when it crosses an interstellar disturbance. 
These authors also use a 3D magnetic field structure but are more
interested in the X-ray features than in the details of the shock
structure. The latter is the topic of this work.

Hot stars are not alone in developing shock structures. Cool F-, G-, K-, and
even M-stars are driving supersonic winds and may show bow shocks if
the star's velocity with respect to the ambient interstellar medium
(ISM) is sufficiently high. Some of these structures of nearby stars
can be observed in Ly-\textalpha\
lines that are produced when neutral hydrogen atoms are slowed
down upon entering the shock region
\citep[see, e.g.,][]{Wood-etal-2007, Linsky-Wood-2014}.

In the literature a couple of 2D HD simulations exist, for example, \citet{Brighenti-DErcole-1997}, \citet{Comeron-Kaper-1998} and  lately \citet[][see also the references therein concerning the HD flows]{Green-etal-2019}. The problem with the HD astrosphere models is always that the tangential discontinuity, the astropause (AP), is notoriously unstable. For a detailed discussion see Section 2.1.

In a series of recent papers
\citep{Scherer-etal-2015a, Scherer-etal-2016a, Scherer-etal-2016c}, we have
modeled and discussed the astrosphere around \LC and its influence on the
modulation of cosmic rays, albeit ignoring magnetic fields and their effects.
We have now
improved the modeling by including both a Parker-like stellar wind magnetic field
at the inner boundary and an interstellar magnetic field beyond the AP. While a few aspects, including a comparison to observations of
astrospheres around cool stars (such as the Sun), can be found in
\citet{Scherer-etal-2016c}, here we continue the discussion using an MHD
model of \LC and compare it with heliosphere-like astrospheres and a wind
bubble.

The paper is structured as follows:
In Section~2, we discuss the models used and the resulting MHD structures.
We compare the O-star model efforts with those  performed for the heliosphere in Section~3,
where we also show and discuss the motion of the bow shock, especially for the
bubble models, and conclude with a summary in Section~4.

\section{General features of the model}
\subsection{The input parameter and the numerical model}

\begin{table*}
  \centering
  \caption{Initial values for the stellar wind (SW) are given at the
    inner boundary and those for the ISM at ``$\infty$''
    (or rather at the outer boundary of the computational
    volume). Note that some models start at different inner boundaries
    (see first row).  The cooling function
    (using solar abundances) is described in
    \citet{Schure-etal-2009} and the heating function in
    \citet{Kosinski-Hanasz-2006}.  }
  \label{tab:c1}
  \begin{tabular}{llllrrr}\toprule
    Region & parameter   & symbol  & unit &\multicolumn{3}{c}{Model} \\ \cmidrule(r{0.5em}l{0.5em}){5-7}
    &&       && \LC  & heliosphere & Wind bubble \\
    \midrule
    -- & inner boundary      & $r_0$ & AU         & 6000 & 1 & 6000 \\
    SW   & temperature  & $T\sw$ & K           & 1000  & 73640 & $10^5$ \\
    SW   & speed        & $v\sw$ & km s$^{-1}$ & 2500 &   375 & 1500 \\
    SW   & angular frequency & $\Omega\sw$ & \textmu Hz & 17.7 & 2.7 & -- \\
    SW   & number density & $n\sw$ & cm$^{-3}$   & 4.1   &     7 &  1 \\
    SW   & magnetic field strength & $B\sw$ & \textmu G & 0.67 & 40  & 100 \\
    \midrule
    ISM  & temperature   &$T\ism$& K          & 10\,000 & 6530  &  1500 \\
    ISM  & speed         &$v\ism$& km s$^{-1}$   & 80  & 26.4  & 0\\
    ISM  & latitude of inflow &$\vartheta_{v,{\rm ism}}$ && 90$\deg$ & 0$\deg$ & -- \\
    ISM  & longitude of inflow &$\varphi_{v,{\rm ism}}$  && 180$\deg$ & 180$\deg$   & --  \\
    ISM  & number density &$n\ism$& cm$^{-3}$  & 11     &  1 & 1.7 \\
    ISM  & magnetic field strength &$B\ism$& \textmu G & 10 & 3 & 0 \\
    ISM  & magnetic field latitude &$\vartheta_{B,{\rm ism}}$ && 45$\deg$ &150$\deg$  & --\\
    ISM  & magnetic field longitude &$\varphi_{B,{\rm ism}}$ && 45$\deg$ & 30$\deg$  & --\\
    -- & cooling function && & Schure & -- & Schure \\ 
    -- & heating function &&& Kosi{\'n}ski \& Hanasz & --& Kosi{\'n}ski \& Hanasz\\\bottomrule
  \end{tabular}
\end{table*}

\begin{figure*}
\includegraphics[width=0.475\textwidth]{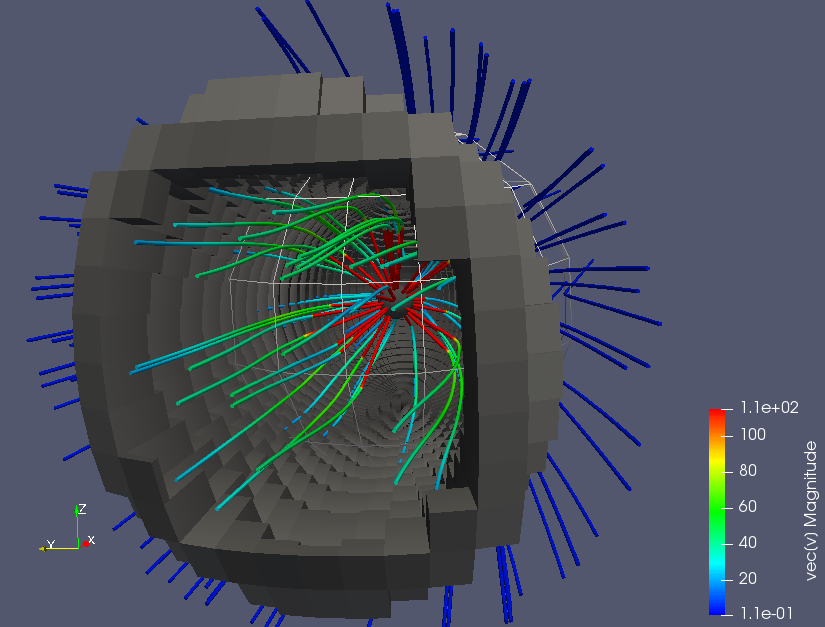}\hfill
\includegraphics[width=0.475\textwidth]{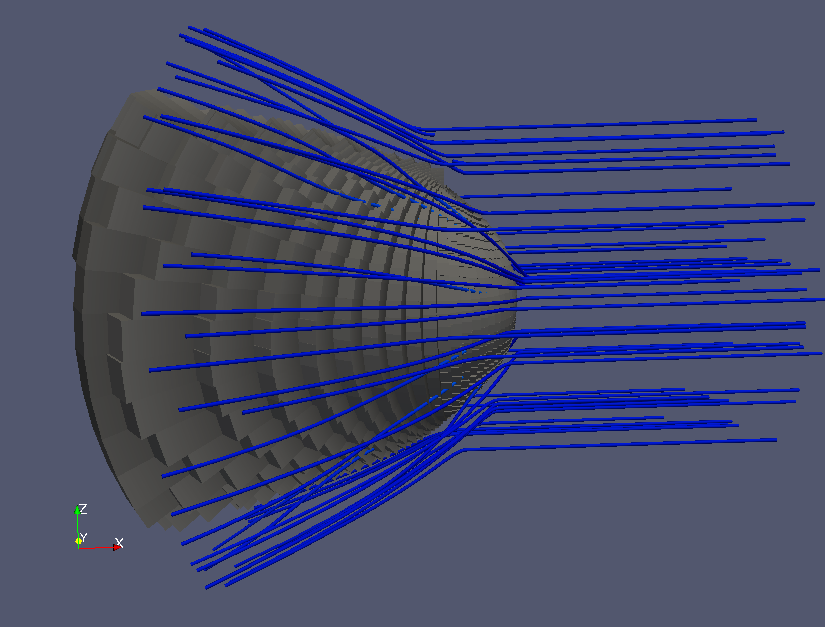}\\[0.25cm]
\includegraphics[width=0.475\textwidth]{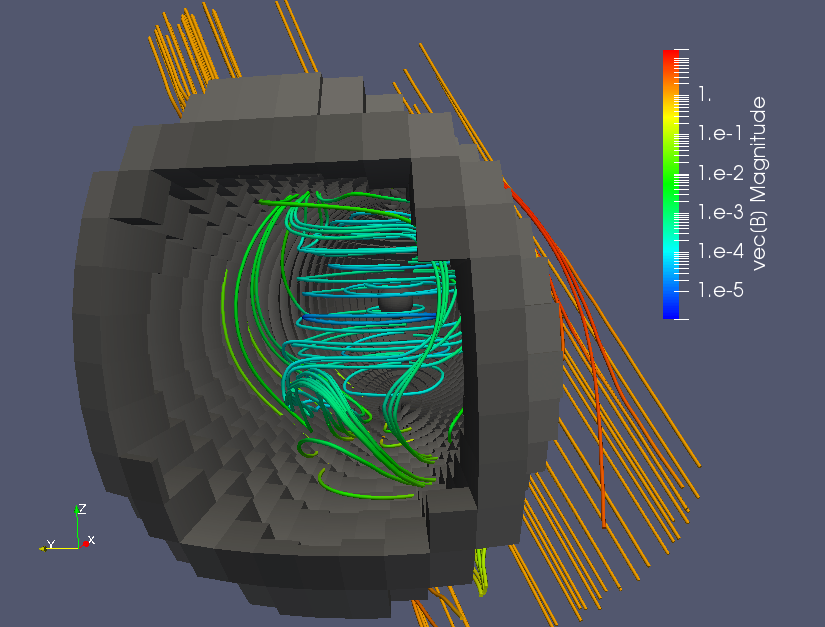}\hfill
\includegraphics[width=0.475\textwidth]{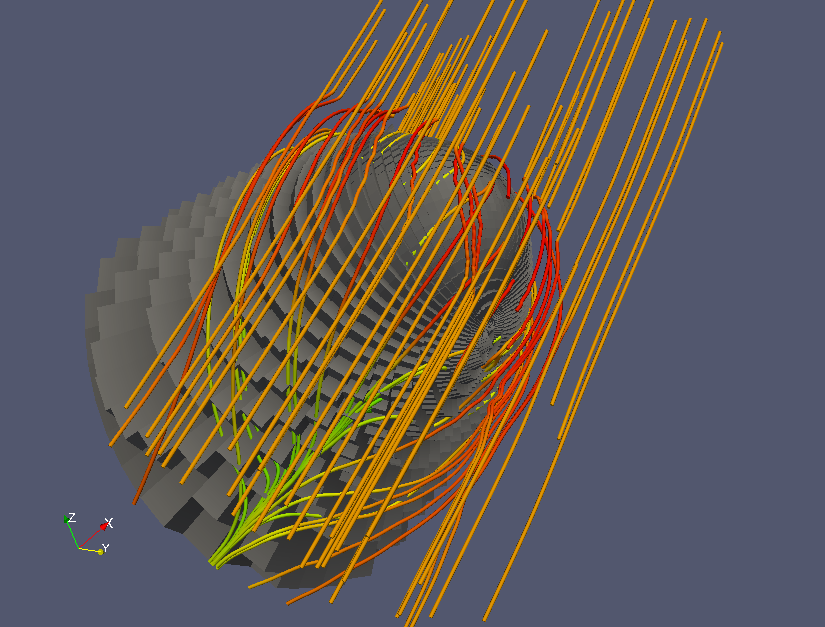}
\caption{Perspective renderings showing the wrapping of the flow lines
  (upper panels) and magnetic
  field (lower panels) around the \LC astropause. The left
  panels depict a view along the tail direction towards the central
  star, which shows the outflow in this direction and the winding of
  the inner field, while in the right panel the
  interstellar flow lines and magnetic field lines are visualized. The
  blocky surface indicates the AP, identified via its characteristic
  jump in number density. The blocks represent individual computational cells.
}
\label{fig:1}
\end{figure*}

We use the 3D finite-volume MHD code \textsc{Cronos} \citep{Kissmann-etal-2018}
based on a Riemann solver to perform simulations on a star-centered spherical grid with a resolution of
$N_r \times N_\vartheta \times N_\varphi = 990 \times 30 \times 60$
cells within a sphere of $r \in [0.03, 10]$~pc and full $4\pi$ solid angle coverage. In this study, three different models will be investigated:
%Table~\ref{tab:c1} lists the physical parameters which characterize the
%three different models:
The astrosphere of \LC, the heliosphere, and a wind bubble. The corresponding physical parameters are listed in Table~\ref{tab:c1}. While in the case of the \LC and the heliosphere models the central
star has a non-zero relative velocity with respect to the
ambient ISM, the wind bubble is at rest. 
As an example, Fig.~\ref{fig:1} shows the magnetic field lines and the flow lines of \LC at a time
$t=1$\,Myr, together with the AP indicated by a contour of its number density. As can be
seen, the magnetic field lines and flow lines are neither parallel nor perpendicular to each other, implying that the shock is neither parallel nor perpendicular.

  The employed angular resolution of
  $6^{\circ}$ per cell
    in both $\vartheta$
    and $\varphi$
    may seem relatively coarse but is justified by noting that
    the ensuing analysis will mostly focus on the shock structure along
    the inflow axis. As will be discussed later, deviations from
    axial symmetry can be expected, and are indeed found to be rather
    small.

 We used in the following the cooling function tested by \citet{Schure-etal-2009} for solar metalicities, and tested also (most of) the cooling functions given by \citet{Sutherland-Dopita-1993}, \citet{Mellema-Lundqvist-2002}, and \citet{Gnat-Ferland-2012}. We did not find any differences to the model presented here. Thus, the conclusion is that for strong interstellar fields the cooling does not play a role for the dynamics because of the magnetic pressure. Nevertheless, for the observational aspects it is important. Thus, the flow can be treated as an ideal MHD scenario and one can apply the Rankine-Hugoniot relations to the TS and BS (see below).

\subsubsection{The stability of the astropause and shocks in (M)HD simulations}

When there exist a velocity shear in a fluid, the Kelvin-Helmholtz
(KH) instability can occur; if two fluids with different densities are
separated by and interface they may be effected by the Rayleigh-Taylor
(RT) instability
\citep[for the HD case see, for example
][]{Landau-Lifshitz-1987, Biskamp-2008}, and
\citep[fort the MHD case][]{Landau-Lifshitz-1984,Biskamp-2008}. Both
instabilities can appear at the AP: In the case of no relative motion
the RT with its characteristic blobs may show up, while in a
counterflow configuration (that is with a relative velocity) the KH
may operate.  These two instabilities 
may both be present in HD simulations, where they grow and may lead to
very large structures, as can be seen in the 2D HD simulations
\citet{Brighenti-DErcole-1997}, \citet{Comeron-Kaper-1998} and lately
\citet{Green-etal-2019}, where by close inspection of their figures a
BS can be found in front of some of the perturbed AP.  These
perturbations at the AP are so large that they influence the entire
integration area, for a detailed discussion for the heliosphere see
\citet{Wang-Belcher-1998}. In a counterflow configuration, the
relevant perturbation is of KH-type, which also does appear at the AP
but is transported into the downwind direction, and usually does not
interfere with the BS or AP. In a 3D HD simulations by
\citet{Reyes-Iturbide-etal-2019} these features can also be found.
  
A detailed analysis of the KH in the case of the heliosphere is given
by \citet{Ruderman-Brevdo-2006}, where the authors show that the
magnetic field stabilizes the astropause. The feature of a smooth
astropause is mainly seen in heliospheric simulations
\citep[for example][]{Pogorelov-etal-2017} and in the 3D MHD
astrosphere simulations by \citet{Katushkina-etal-2018} and
\citet{Gvaramadze-etal-2018}.  Most of the instabilities are discussed
in detail for the heliosphere, a special astrosphere.

We also want to emphasis here that in HD the Bernoulli law holds,
which states that the total pressure is constant along a
streamline. This is not the case in MHD, because in momentum and
energy equations there exist additional terms connected to the 3D
character of the magnetic field (.i.e.\ $\vec{B}\otimes\vec{B}$
and $\vec{B}(\vec{v}\cdot\vec{B})$, respectively).

In our models, the spatial cell size in $r$-direction
is in the order of $ 0.01$\,pc
with an angular resolution of about 5$^{o}$.
However, to include instabilities like, for example, the KH
instability, this resolution is not sufficient. The length scale
$\delta$
for the KH can be estimated based on a pure MHD flow using linear
perturbation theory by
\mbox{$k \delta \approx 1$}
  \citep[see][]{Biskamp-2008}, where $k$
  is the wavenumber of the maximal growth rate of the KH. Assuming
  that $\delta$
  is of the order of a few proton gyroradii $R_{\mathrm{g}}$,
  in particular $R_{\mathrm{g}}=10^{-7}$\,pc
  in the outer astrosheath and much smaller in the inner one, the
  required resolution is around $10^{-7}$\,pc.
  This, however, is far beyond current computational resources and,
  therefore, cannot be taken into account. This, furthermore, also
  applies to other processes like reconnection that requires a
  resolution of a few gyroradii.
\subsubsection{The shock and astropause distances}
For the cases based on HD modeling, in which the Bernoulli equation holds, the distance $r\TS$ to the termination shock (TS) can be calculated as
\begin{equation}
  \label{eq:0}
  r\TS =
  r_{0} \sqrt{\frac{\rho_{0, {\rm sw}} \ v\sw^{2}}{\rho\ism \, v\ism^{2}}} =
  \sqrt{\frac{\dot{M} v\sw}{4\pi \, \rho\ism \, v\ism^{2} }}\ ,
\end{equation}
using either the stellar mass loss rate $\dot{M}$ or the stellar wind mass density
$\rho_{0, {\rm sw}}$ at a specific reference distance $r_0$. In this study $r_0$ is set as the inner radial boundary. Note that the latter expression of Eq.~(\ref{eq:0}) is know as the Wilkin formula \citep{Wilkin-2000}, while the first one was first noted by
\citet{Parker-1958}. For a more
detailed discussion see also \citet{Scherer-etal-2016a}.
Although this being the termination shock distance, other authors \citep[among others]{delValle-Pohl-2018, Katushkina-etal-2017} call the above distance the standoff distance or the bow shock distance \citep{Meyer-etal-2016}. The reason why only the TS distance can be determined is the fact that in ideal HD between the TS and the BS  Bernoulli's law holds, while in the downstream region of TS and BS the ram ($\rho v^2$/2) and thermal pressure ($P$) can be calculated via the Rankine-Hugoniot equations. 
The TS distance (see Eq.~(\ref{eq:0})) will be used in the following to normalize the
distances.

\begin{figure*}
    \centering
    \includegraphics[width=0.49\textwidth]{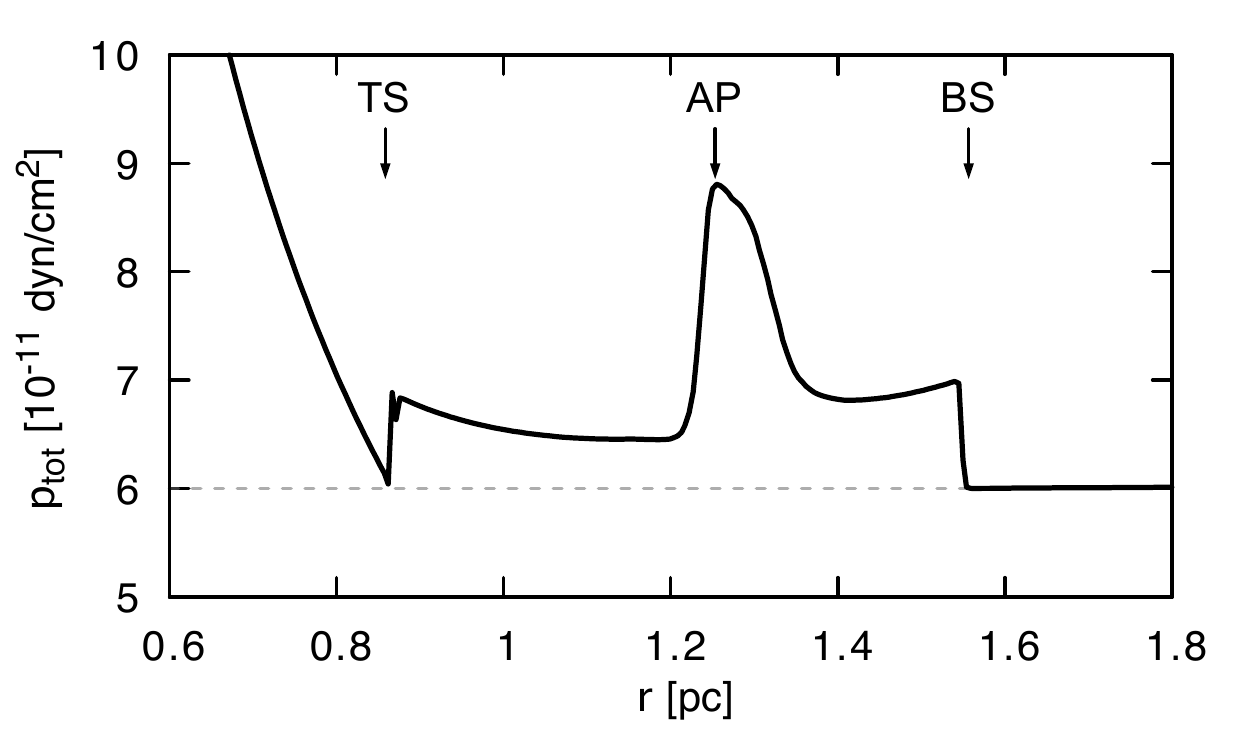}
    \includegraphics[width=0.49\textwidth]{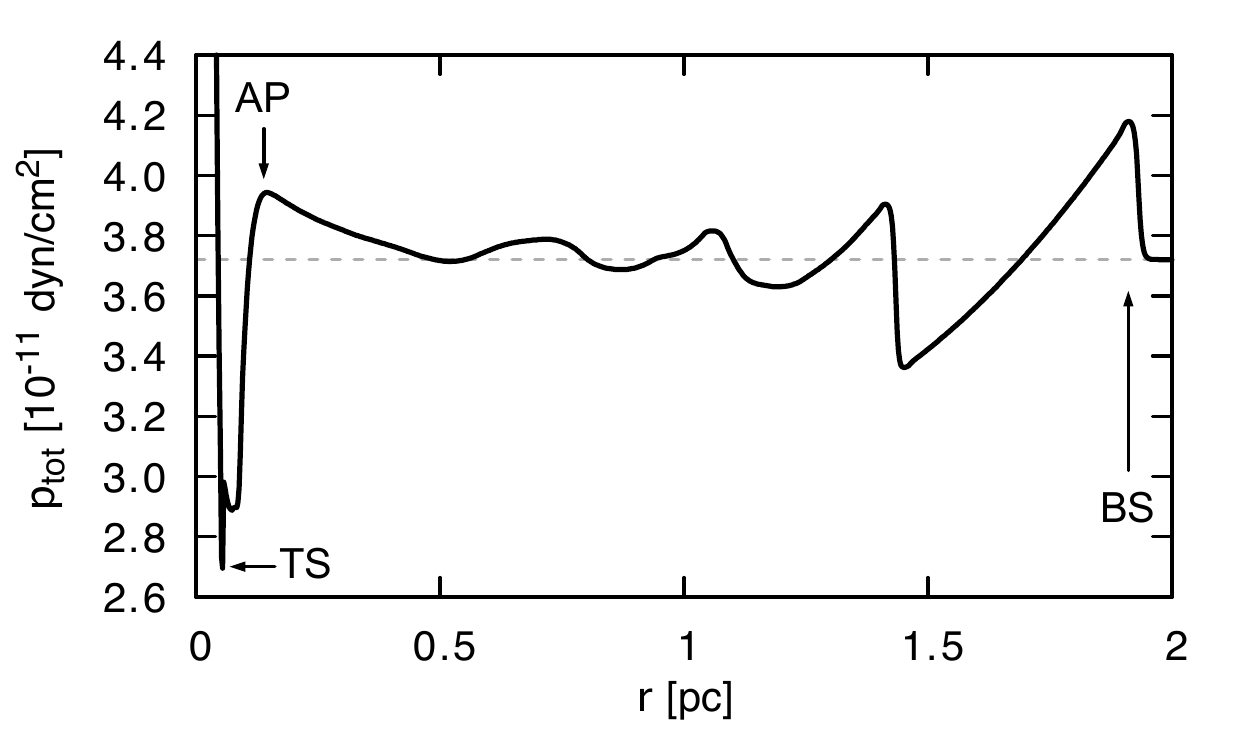}
    \caption{The total pressure along the inflow line. In the left panel the total pressure for \LC is presented, while in the right panel that for VPeg 374. It can be seen that the total pressure is not conserved, nor that it must have the same value at the BS and TS.}
    \label{fig:bernoulli}
\end{figure*}

Because for O-stars the magnetic field pressure is usually
much lower than the ram pressure, this approach is a useful approximation
for MHD supersonic and superfast magnetosonic scenarios which also include
heating and cooling. However, there is nothing like a Bernoulli law in ideal MHD because of the terms $\vec{B}\odot\vec{B}$ and $(\vec{B}\cdot\vec{v})\vec{B}$ in the momentum, respectively in the energy equation. Figure~\ref{fig:bernoulli} shows the total pressure $\rho v^2/2 + P +B^2/(8\pi)$ for \LC (left panel) and V374Peg (right panel) along a line through the star and parallel to inflow velocity at infinity, which is a streamline for the HD, but not for the MHD case. This is clearly seen in the figure, where for \LC the total pressure at the TS and BS, by chance, is the same but definitely not for V374Peg, where different flow lines are crossed due to the asymmetry. Thus, across the flow lines, the pressure is not constant, which is indicated by the different peaks visible between the AP and the BS. Also, in the downstream regions between the shocks, the total pressure in both cases is not constant. Thus, in the case of MHD model efforts, Eq.~(\ref{eq:0}) gives only an approximation of the TS distance.

\begin{figure*}
    \centering
 \includegraphics[width=\textwidth]{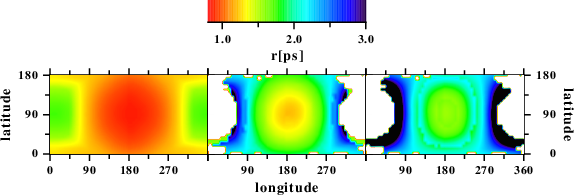}
    \caption{The distances of the TS (left panel), the AP (middle panel) and the BS (right panel). Note the color bars indicating the distances are different for each panel. In black areas around the colored feature the distances are not defined.} 
    \label{fig:ts}
\end{figure*}

Further, we determined the TS, AP, and BS distances of \LC (see Fig.~\ref{fig:ts}) for the six different directions listed in Table~\ref{tab:d1}: the nose- and tail-ward directions, the east- and westward directions, and those over the two poles. The jump in the thermal pressure determines the TS and BS. The latter is much larger than that of the density or speed, while the AP distance is given by the minimum of the sonic Mach number. We also present the upwind and downwind distance of the TS using Eq.~(\ref{eq:0}). In principle (for the MHD case), the magnetosonic Mach numbers are a possible choice, but it is not guaranteed that they jump from above to below one as in the HD case. Additionally, in the case of (M)HD, the so-called weak solution can occur in the flanks, where the Mach numbers can be larger than one \citep[see][]{Scherer-etal-2016a}.   

As can be seen, in the case of \LC the analytically estimated upwind TS distance, and the one determined from the model are almost identical. 
We also give the asymmetry between the two opposite directions (i.e., \ upwind/downwind, east/west and north/south) as well as the MHD shock type. Only for the TS in the upwind direction an analytic calculation using the HD case can be given. To calculate the TS distance analytically in the tail direction, we assumed an asymmetry of two, which is used in Tables~\ref{tab:ts} to~\ref{tab:bs}. For convenience, the projected distances are shown in Fig.~\ref{fig:ts}, reflecting that only the TS is defined for all directions while the AP and BS distances are not (the white areas in Fig.~\ref{fig:ts}).

\begin{table}
 \caption{\label{tab:d1}Distances to the TS, AP and BS. In 
 in the sixth column the ratio of the $r_{\mathrm{d}}/r_{\mathrm{u}}, r_{\mathrm{E}}/r_{\mathrm{W}}$  and $r_{\mathrm{S}}/r_{\mathrm{N}}$ are given. The MHD shock types are also indicated, where "fast" is a fast shock, TD is the tangential discontinuity, and "no" means that there is no shock.}
 \begin{tabular}{lrrlp{1cm}p{1cm}}\toprule
 Distance & analytic& Model & units& Shock type&Asym\-metry\\
 \midrule
 $r_{\mathrm{u}}$ (TS) & 0.87& 0.87& pc& fast& 2.08\\
 $r_{\mathrm{d}}$ (TS) & 1.74& 1.81& pc& fast& --\\
 $r_{\mathrm{W}}$ (TS) & --& 1.07& pc& fast& 1.04\\
 $r_{\mathrm{E}}$ (TS) & --& 1.11& pc& fast& --\\
 $r_{\mathrm{N}}$ (TS) & --& 1.19& pc& fast& 1.00\\
 $r_{\mathrm{S}}$ (TS) & --& 1.20& pc& fast& --\\
 \midrule
 $r_{\mathrm{u}}$ (AP) & --& 1.25& pc& TD& n.a.\\
 $r_{\mathrm{d}}$ (AP) & --& 2.95& pc& TD& --\\
 $r_{\mathrm{W}}$ (AP) & --& 1.85& pc& TD& 1.07\\
 $r_{\mathrm{E}}$ (AP) & --& 1.98& pc& TD& --\\
 $r_{\mathrm{N}}$ (AP) & --& 2.01& pc& TD& 1.01\\
 $r_{\mathrm{S}}$ (AP) & --& 2.03& pc& TD& --\\
 \midrule
 $r_{\mathrm{u}}$ (BS) & --& 1.86& pc& fast & n.a.\\
 $r_{\mathrm{d}}$ (BS) & --& 4.66& pc& fast& --\\
 $r_{\mathrm{W}}$ (BS) & --& --& pc& no & n.a.\\
 $r_{\mathrm{E}}$ (BS) & --& --& pc& no & --\\
 $r_{\mathrm{N}}$ (BS) & --& --& pc& no & n.a.\\
 $r_{\mathrm{S}}$ (BS) & --& --& pc& no & --\\ \bottomrule
 \end{tabular}
\end{table}

\subsubsection{The analytical Rankine-Hugoniot and the modeled shock relations}
The MHD Rankine-Hugoniot relations also allow an analytic description of the relevant parameters at the TS and BS under the assumption that we know the shock normal vector $\vec{n}$. The relevant equations can be found in any textbook on MHD \citep[e.g., \ ][]{Goedbloed-etal-2010} and are given in Appendix A in the rest frame of the shocks. We assume here that the shock normal $\vec{n}$ is always directed along the $x$-axis. Some of the scalar upstream and downstream parameters for the TS and the BS are shown in Table~\ref{tab:ts} and Table~\ref{tab:bs}, respectively. Furthermore,  Table~\ref{tab:vec} gives the magnetic field and velocities (upstream and downstream) in a Cartesian coordinate system and in the frame constructed by the normal vector and the tangential ones. The tables are organized in such a way that we give first the boundary conditions (inner one for the TS and the outer one for the BS), then the parameters calculated with the assumption that the shocks are hydrodynamical; in the next row the analytically estimated MHD values are given, followed by those derived from the numerical model. The tables also list the upstream and downstream values. It can be seen that the TS is almost a hydrodynamical shock because of the huge Alf\'enic Mach numbers compared to the sonic ones, whereas the BS is a fast, genuine MHD shock.  

\begin{table*}
\caption{\label{tab:ts}Stellar wind parameters derived from the Rankine-Hugoniot 
 relations for ideal HD and MHD as well as for the numerical model (see Appendix A).
 The second row gives the parameters at the inner boundary, the
 third to fifth are the parameters in front of the TS for the analytic HD and MHD
 case and for the numerical model, respectively. The rows six to eight are
 the parameters beyond the shock. For the analytic cases, the normal vector is 
 directed along the $x$-axis, while for the numerical case it is calculated via the
 coplanarity theorem. The parameters $c,M,n,v,T,P,R,s$ are the sound speed, the sonic Mach number, number density, speed, temperature, pressure, ram pressure, and compression ratio, respectively.}
 
 \scalebox{0.8}{
  \begin{tabular}{lrrrrrrrl}\toprule
 Parameter & $r_{\min}$  &\multicolumn{3}{c}{upstream} 
 & \multicolumn{3}{c}{downstream} &units\\ \cmidrule(r{0.2em}l{0.2em}){3-5}\cmidrule(r{0.2em}l{0.2em}){6-8}
 &&HD Analytic& MHD Analytic & Model & HD Analytic & MHD Analytic & Model\\
 \midrule
$c_{\mathrm{n,sw}}$& 11.73& 1.86& 1.86& 0.61& 1173.94& 1173.94& 1730.15 & km/s\\
$c_{\mathrm{t,sw}}$& 11.73& 1.16& 1.16& 0.38& 1173.94& 1173.94& 1670.61 & km/s\\
$M_{\mathrm{n,sw}}$& 179.04& 1131.94& 1131.94& 4096.87& 0.45& 0.45& 0.34 & \\
$M_{\mathrm{t,sw}}$& 179.04& 1803.77& 1803.77& 6607.85& 0.45& 0.45& 0.42 & \\
$n_{\mathrm{n,sw}}$& 4.08& $1.61 \cdot 10^{-02}$& $1.61 \cdot 10^{-02}$& $1.17 \cdot 10^{-02}$& $6.46 \cdot 10^{-02}$& $6.46 \cdot 10^{-02}$& $4.59 \cdot 10^{-02}$
 & \#/cm$^{3}$\\
$n_{\mathrm{t,sw}}$& 4.08& $3.99 \cdot 10^{-03}$& $3.99 \cdot 10^{-03}$& $2.82 \cdot 10^{-03}$& $1.60 \cdot 10^{-02}$& $1.60 \cdot 10^{-02}$& $1.05 \cdot 10^{-02}$
 & \#/cm$^{3}$\\
$v_{\mathrm{n,sw}}$& 2100.00& 2100.00& 2100.00& 2500.02& 525.00& 525.00& 587.78 & km/s\\
$v_{\mathrm{t,sw}}$& 2100.00& 2100.00& 2100.00& 2500.02& 525.00& 525.00& 706.53 & km/s\\
$T_{\mathrm{n,sw}}$& $1.00 \cdot 10^{4}$& 250.19& 250.19& 27.07& $1.00 \cdot 10^{8}$& $1.00 \cdot 10^{8}$& $2.18 \cdot 10^{8}$
 & K\\
$T_{\mathrm{t,sw}}$& $1.00 \cdot 10^{4}$& 98.53& 98.53& 10.41& $1.00 \cdot 10^{8}$& $1.00 \cdot 10^{8}$& $2.18 \cdot 10^{8}$
 & K\\
$P_{\mathrm{n,sw}}$& $8.45 \cdot 10^{-12}$& $8.37 \cdot 10^{-16}$& $8.37 \cdot 10^{-16}$& $1.32 \cdot 10^{-16}$& $1.34 \cdot 10^{-09}$& $8.93 \cdot 10^{-10}$& $1.38 \cdot 10^{-09}$
 & dyne / cm$^{2}$\\
$P_{\mathrm{t,sw}}$& $8.45 \cdot 10^{-12}$& $8.14 \cdot 10^{-17}$& $8.14 \cdot 10^{-17}$& $1.21 \cdot 10^{-17}$& $3.31 \cdot 10^{-10}$& $2.21 \cdot 10^{-10}$& $2.94 \cdot 10^{-10}$
 & dyne / cm$^{2}$\\
$R_{\mathrm{n,sw}}$& $1.51 \cdot 10^{-07}$& $5.96 \cdot 10^{-10}$& $5.96 \cdot 10^{-10}$& $6.14 \cdot 10^{-10}$& $1.49 \cdot 10^{-10}$& $1.49 \cdot 10^{-10}$& $1.33 \cdot 10^{-10}$
 & dyne / cm$^{2}$\\
$R_{\mathrm{t,sw}}$& $1.51 \cdot 10^{-07}$& $1.47 \cdot 10^{-10}$& $1.47 \cdot 10^{-10}$& $1.47 \cdot 10^{-10}$& $3.68 \cdot 10^{-11}$& $3.68 \cdot 10^{-11}$& $4.38 \cdot 10^{-11}$
 & dyne / cm$^{2}$\\
$s_{\mathrm{n,sw}}$& --& --& --& --& 4.00& 4.00& 3.90 & \\
$s_{\mathrm{t,sw}}$& --& --& --& --& 4.00& 4.00& 3.72 & \\
 \bottomrule
 \end{tabular}
 }
 \end{table*}
 \begin{table*}
 \caption{\label{tab:bs} Interstellar medium parameters.
 The columns are ordered in the same way as in Table~\ref{tab:ts}.}
  \scalebox{0.8}{
 \begin{tabular}{lrrrrrrrl}\toprule
 Parameter & $r_{\min}$  &\multicolumn{3}{c}{upstream} 
 & \multicolumn{3}{c}{downstream} &units\\ \cmidrule(r{0.2em}l{0.2em}){3-5}\cmidrule(r{0.2em}l{0.2em}){6-8}
 &&HD Analytic& MHD Analytic & Model & HD Analytic & MHD Analytic & Model\\
 \midrule
$c_{\mathrm{n,ism}}$& 11.13& 11.13& 11.13& 11.22& 45.49& 45.13& 55.39 & km/s\\
$M_{\mathrm{n,ism}}$& 7.19& 7.19& 7.19& 6.98& 0.46& 0.48& 0.41 & \\
$n_{\mathrm{n,ism}}$& 11.00& 11.00& 11.00& 11.31& 41.59& 40.67& 39.70 & \#/cm$^{3}$\\
$v_{\mathrm{n,ism}}$& 80.00& 80.00& 80.00& 78.38& 21.16& 21.64& 22.45 & km/s\\
$T_{\mathrm{n,ism}}$& 8999.24& 9000.00& 9000.00& 9155.51& $1.50 \cdot 10^{5}$& $1.48 \cdot 10^{5}$& $2.23 \cdot 10^{5}$
 & K\\
$P_{\mathrm{n,ism}}$& $4.10 \cdot 10^{-11}$& $1.37 \cdot 10^{-11}$& $1.37 \cdot 10^{-11}$& $4.29 \cdot 10^{-11}$& $8.80 \cdot 10^{-10}$& $8.31 \cdot 10^{-10}$& $1.22 \cdot 10^{-09}$
 & dyne / cm$^{2}$\\
$R_{\mathrm{n,ism}}$& $5.89 \cdot 10^{-10}$& $1.18 \cdot 10^{-09}$& $5.89 \cdot 10^{-10}$& $5.81 \cdot 10^{-10}$& $3.11 \cdot 10^{-10}$& $1.59 \cdot 10^{-10}$& $1.67 \cdot 10^{-10}$
 & dyne / cm$^{2}$\\
$s_{\mathrm{n,ism}}$& --& --& --& --& 3.78& 3.70& 3.51 & \\
 \bottomrule
 \end{tabular}
 }
 \end{table*}
 \begin{table*}
 \caption{\label{tab:vec}The wind magnetic field and velocity vectors
 in the ecliptic at the TS. 
 The indices $1,2$ are upstream and downstream, respectively, and the
 indices $\mathrm{n,t}$ stand for the normal component and the tangential vectors. }
 \scalebox{0.8}{
  \begin{tabular}{lrrrrrrl}\toprule
 Parameter & \multicolumn{3}{c}{MHD Analytic} 
 & \multicolumn{3}{c}{Model} &units\\ \cmidrule(r{0.2em}l{0.2em}){2-4}\cmidrule(r{0.2em}l{0.2em}){5-7}
 \midrule
$\vec{n}$& 1.00& $0$& $0$& 0.88& -0.34& 0.33 & \\
$\vec{B}_{1}$& $6.12 \cdot 10^{-06}$& $2.93 \cdot 10^{-11}$& $-1.29 \cdot 10^{-18}$& $2.38 \cdot 10^{-07}$& $-1.25 \cdot 10^{-08}$& $-1.19 \cdot 10^{-08}$
 & $\si{\micro G}$\\
$B_{\mathrm{n}}$& $6.12 \cdot 10^{-06}$& & & $2.09 \cdot 10^{-07}$& &  & $\si{\micro G}$\\
$\vec{B}_{1,\mathrm{t}}$& $0$& $2.93 \cdot 10^{-11}$& $-1.29 \cdot 10^{-18}$& $5.37 \cdot 10^{-08}$& $5.97 \cdot 10^{-08}$& $-8.09 \cdot 10^{-08}$
 & $\si{\micro G}$\\
$\vec{B}_{2}$& $6.12 \cdot 10^{-06}$& $1.17 \cdot 10^{-10}$& $-5.16 \cdot 10^{-18}$& $2.22 \cdot 10^{-07}$& $-3.04 \cdot 10^{-08}$& $1.24 \cdot 10^{-08}$
 & $\si{\micro G}$\\
$\vec{B}_{2,\mathrm{t}}$& $0$& $1.17 \cdot 10^{-10}$& $-5.16 \cdot 10^{-18}$& $3.76 \cdot 10^{-08}$& $4.18 \cdot 10^{-08}$& $-5.66 \cdot 10^{-08}$
 & $\si{\micro G}$\\
$\vec{v}_{1}$& 2100.00& $0$& $0$& 2493.17& $-130.66$& $-130.84$ & km/s\\
$v_{1,\mathrm{n}}$& 2100.00& & & 2193.89& &  & km/s\\
$\vec{v}_{1,\mathrm{t}}$& $0$& $0$& $0$& 564.32& 624.88& $-853.23$ & km/s\\
$\vec{v}_{2}$& 525.00& $7.53 \cdot 10^{-17}$& $-3.32 \cdot 10^{-24}$& 585.44& $-52.26$& $-4.86$
 & km/s\\
$v_{2,\mathrm{n}}$& 525.00& & & 531.11& &  & km/s\\
$\vec{v}_{2,\mathrm{t}}$& $0$& $7.53 \cdot 10^{-17}$& $-3.32 \cdot 10^{-24}$& 118.49& 130.64& $-179.74$
 & km/s\\ \bottomrule
 \end{tabular}
}
\end{table*}
 
  \begin{table*}
  \caption{\label{tab:vecism}The LISM magnetic field and velocity vectors
 in the ecliptic at the BS. 
 The indices are analogous to those in Table~\ref{tab:vec}.}
 \scalebox{0.8}{
 \begin{tabular}{lrrrrrrl} \toprule
 Parameter & \multicolumn{3}{c}{MHD Analytic} 
 & \multicolumn{3}{c}{Model} &units\\ \cmidrule(r{0.2em}l{0.2em}){2-4}\cmidrule(r{0.2em}l{0.2em}){5-7}
 \midrule
$\vec{n}$& $-1.00$& $0$& $0$& $-1.00$& $-7.66 \cdot 10^{-03}$& $7.06 \cdot 10^{-02}$
 & \\
$\vec{B}_{1}$& $-4.33$& 2.50& 8.66& $-4.33$& 2.50& 8.66 & $\si{\micro G}$\\
$B_{\mathrm{n}}$& 4.33& & & 4.47& &  & $\si{\micro G}$\\
$\vec{B}_{1,\mathrm{t}}$& $0$& 2.50& 8.66& $-0.82$& $-3.97$& $-11.96$ & $\si{\micro G}$\\
$\vec{B}_{2}$& $-4.33$& 9.28& 32.13& $-6.52$& $-10.09$& $-29.95$ & $\si{\micro G}$\\
$\vec{B}_{2,\mathrm{t}}$& $0$& 9.28& 32.13& $-2.06$& $-10.06$& $-30.26$ & $\si{\micro G}$\\
$\vec{v}_{1}$& $-80.00$& $0$& $0$& $-77.96$& $7.28 \cdot 10^{-02}$& 8.08 & km/s\\
$v_{1,\mathrm{n}}$& 80.00& & & 78.33& &  & km/s\\
$\vec{v}_{1,\mathrm{t}}$& $0$& $0$& $0$& 0.18& 0.67& 2.55 & km/s\\
$\vec{v}_{2}$& $-21.63$& 0.16& 0.55& $-21.92$& $-2.41$& 4.22 & km/s\\
$v_{2,\mathrm{n}}$& 21.63& & & 22.18& &  & km/s\\
$\vec{v}_{2,\mathrm{t}}$& $0$& $-0.16$& $-0.55$& 0.20& $-2.24$& 2.65 & km/s\\ \bottomrule
 \end{tabular}
 }
 \end{table*}
 
 With the above values, other parameters, like Alfv\'en speeds, plasma beta, for example, can be easily determined. Due to the analytic considerations, one has to assume a normal vector, which we aligned along the inflow axis (the $x$-axis). Calculating the normal vector via the coplanarity theorem, it will be in general different to the above, simplified assumption. The normal vectors are for both cases displayed in Tables~\ref{tab:vec} and~\ref{tab:vecism}. The difference in the analytic and model normal vector also causes the differences in the numbers in Tables~\ref{tab:ts} to~\ref{tab:vecism}. Thus, to get a rough idea of the astrosphere in mind, one can use the discussed analytic approach but should be aware that a 3D model can differ substantially in all directions, especially in those not aligned with the inflow vector. 

\subsection{Observables}
\begin{figure*}
  \includegraphics[scale=0.38]{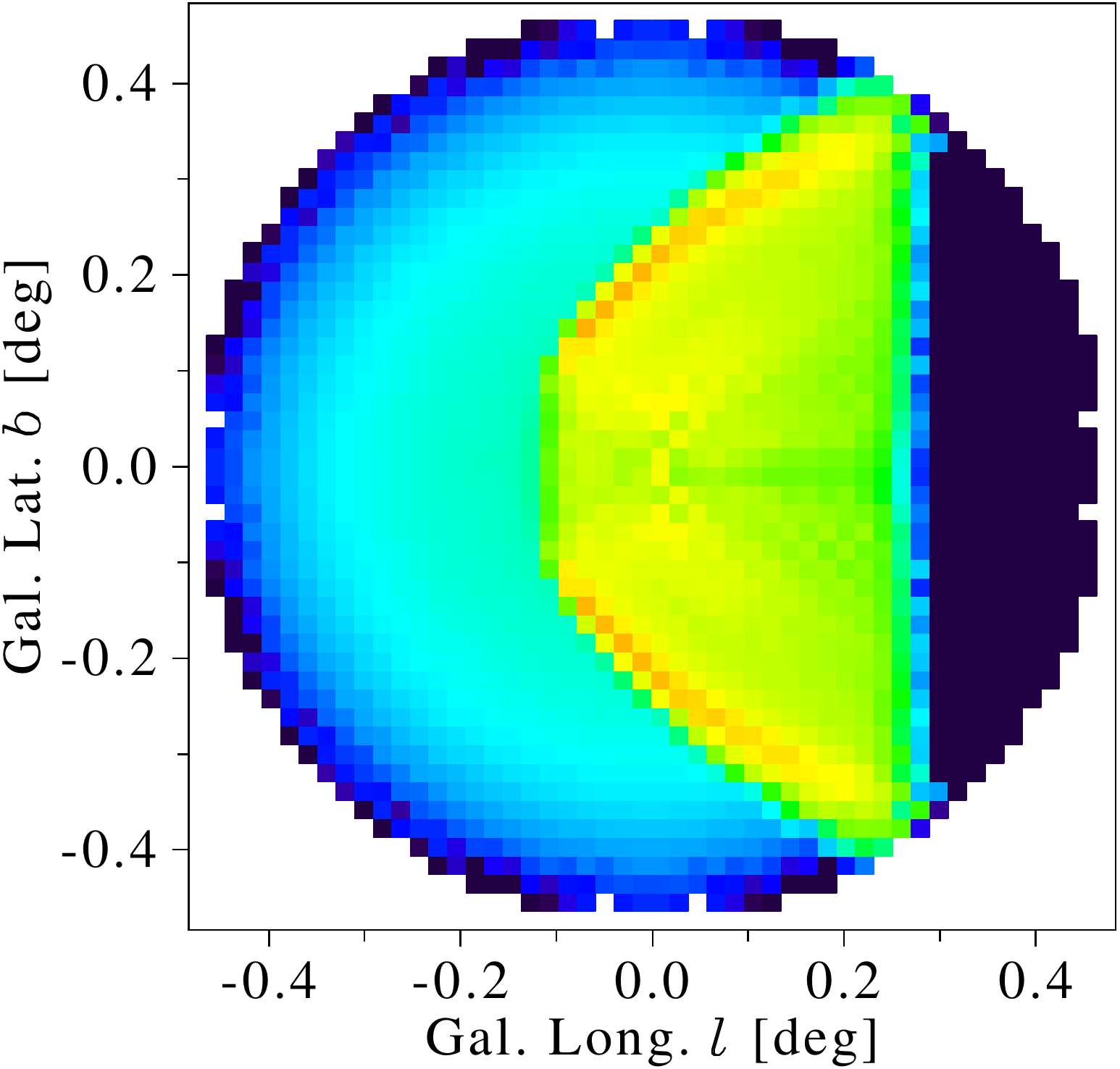}\hfill
  \includegraphics[scale=0.38]{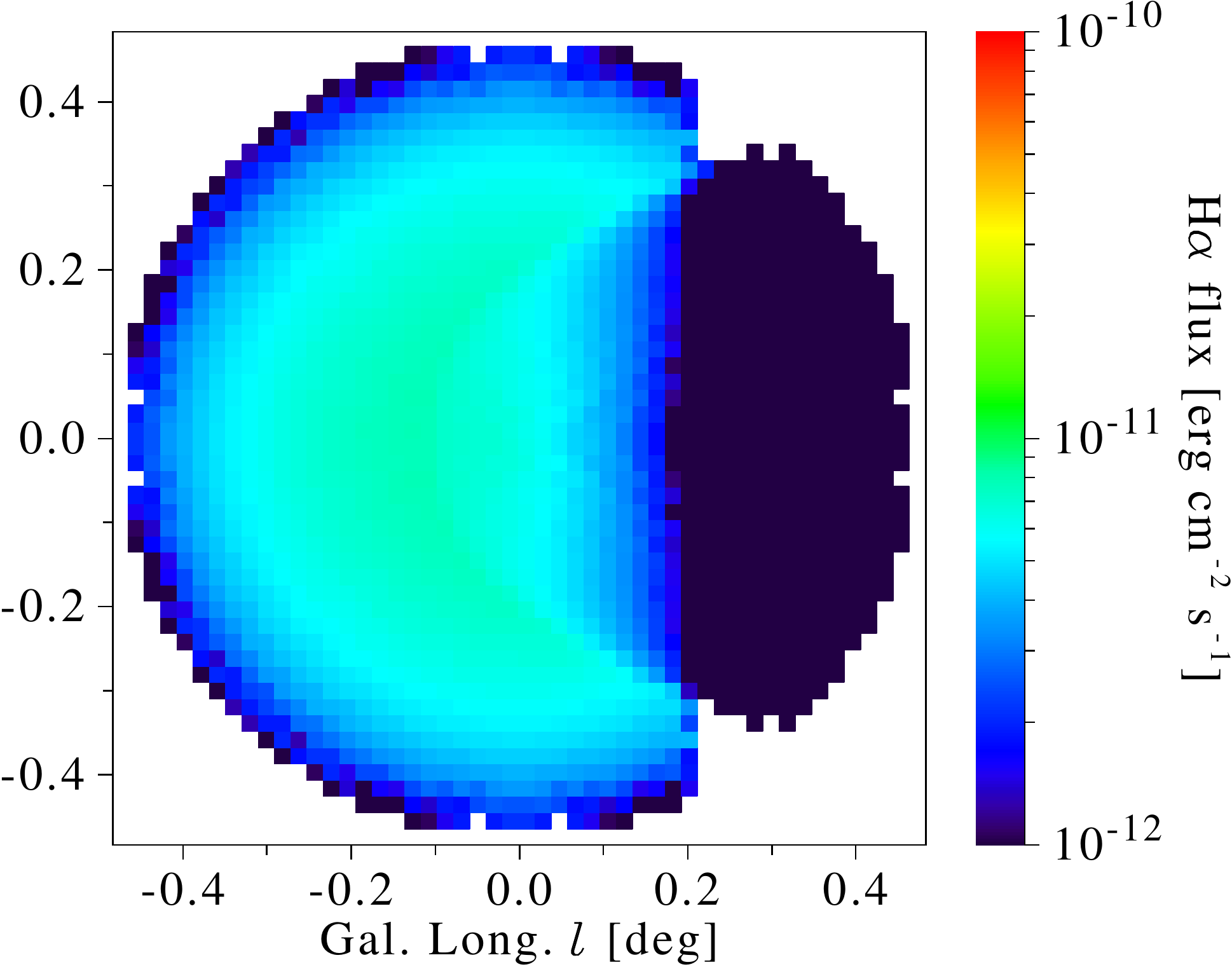}
  \caption{ \label{fig:0}
    The left panel shows the full line-of-sight (LOS) integration of the
    \Ha flux through the
    astrosphere, while in the right panel all the contributions from
    high-density ($n>12\,\text{cm}^{-3}$) areas are excluded. The high-density area (in the left panel)
    can be identified as a region close to the astropause, where the peak in
    density appears. White notches appear in the right panel where all model cells were excluded in the respective lines of sight due to the density criterion.}
\end{figure*}

Unfortunately, the number density at the TS and beyond up to
the AP is very low (around $10^{-3}$ to $10^{-2}\,\si{cm}^{-3}$), thus even with a very hot inner astrosheath (i.e.\
the region between TS and AP) the production of X-rays or \Ha
is strongly suppressed. The number density beyond the AP, however, is much higher, especially due to cooling effects
\citep[see][]{Arthur-2012, Mackey-etal-2014b, Scherer-etal-2016a}.  These regions
are therefore much more likely to be seen in \Ha or other observational
channels \citep{Mackey-etal-2015}. The left panel of Fig.~\ref{fig:0} shows the line-of-sight (LOS) integration of the computed \Ha glow through the entire astrosphere, while the right panel displays the outer astrosheath excluding the high-density region ($n>12$\,cm$^{-3}$). We, therefore, conclude that the \Ha glow mainly comes from the region in which cooling increases the density of the interstellar material. Thus, the total pressure remains constant along a streamline, but because the ideal gas equation implies \mbox{$P \propto n \, T$}, the density $n$ must
increase when the temperature $T$ decreases in order to keep the pressure $P$ constant.
Furthermore, it can be seen that the contribution from other
regions close to the BS, the TS, or the AP are negligible.
Indeed, the right panel of Fig.~\ref{fig:0}
shows only the background, for example, the recombination rate is suppressed by one order of magnitude compared to what is shown in the left panel, where the outer astrosheath is the main contributor.
Due to the scale invariance inherent in the ideal MHD equations,
models can often be transferred between different parameter regimes by
rescaling, so the parameter space of different models to be
investigated is greatly
reduced. Unfortunately, this scale invariance is broken by the cooling
functions, which act mainly in the outer astrosheath, i.e., \ the region between
AP and BS, and cause the extent of this region to shrink.
Because the stellar wind pressure at the astropause still needs to be
balanced, this leads, again, via the ideal gas law, to an increase of the
number density in the outer astrosheath close to the AP. The outer
AP thus shrinks because close to the BS, the material is cooled, and
the pressure is reduced. Finally, the temperature reaches its interstellar equilibrium value, and the pressure can only be increased through an increase
in density. The presence of a magnetic field complicates this
process because then the respective sums of magnetic and thermal pressures,
rather than thermal pressures alone, have to be balanced at the AP,
which in fact is a tangential discontinuity and not a contact discontinuity. For the latter, see, for example, \citet{Goedbloed-etal-2010}. 
Moreover, it should be noted that the fast magnetosonic speed should be lower
than the fluid speed in order to maintain a bow shock structure. Otherwise, the bow shock vanishes, and a bow wave may appear
\citep[see, e.g.,][]{Zank-etal-2013}.

Looking at \Ha images of stars \citep[see][for
  \LC]{Scherer-etal-2016a}, no instabilities can be seen. There are two possible explanations: (1) the observation resolution of current instruments is too small to resolve such features, or (2) the emission does not appear in the astropause, but in front of it. The latter is clearly supported by our previous findings \citep[see][]{Scherer-etal-2016a}, where the number density of the protons in front of the astropause increases dramatically. As a consequence, a ``hydrogen wall'' in front of the astropause will appear, similar to that of the heliopause. However,
  the physical processes are different: while at the heliopause charge exchange processes with interstellar neutral hydrogen play a crucial role, in astrospheres, recombination is the most relevant physical process. The hydrogen wall can be seen when comparing the left and right panels of Fig.~\ref{fig:0}. In the latter case, instabilities like the KH
  cannot be seen in H\textalpha. The KH can only appear at the astropause, which is a tangential discontinuity with different parallel (to the astropause) speeds on both sides.  

As discussed in the following, due to the additional magnetosonic waves, the MHD structure is
much more complex than the one following from pure HD.
\section{Detailed MHD structures}
\subsection{\LC-like astrospheres}
\subsubsection{The initial magnetic field}
The stellar-wind magnetic field $B\sw$ at the inner boundary
\mbox{($r_0 \approx 0.03$\,pc)}, which is assumed to be frozen-in at
that boundary is chosen in such a way that it corresponds to a 100 times stronger magnetic field at $1\,\si{AU}$ compared to that of the solar wind,
$B_{\odot, {\rm w}}$. This is based on observations of stellar surface magnetic
fields $B_{\rm ss}$ \citep[e.g.][]{Peri-etal-2015} 
and \citet{Walder-etal-2012} who gave a review of stellar magnetic fields, which are often multipoles or fossil fields. At large distances the multipoles behave like a dipole, which we adopt here. Because only stellar surface fields are observed, we
 compare those
with the solar surface magnetic field
$B_{\odot. {\rm s}}$.
If the latter fields differ, we assume that the stellar wind
magnetic fields differ the same way, i.e.
\begin{equation}
  \left. \frac{B_{\rm sw}}{ B_{\odot,{\rm w}}} \right|_{1\,\si{AU}}
  = \frac{B_{\rm ss}}{B_{\odot,{\rm s}}} .
\end{equation}
Beyond the last critical point of the stellar wind \citep{Lamers-Cassinelli-1999}, we assume that the magnetic field is passively advected in the flow, and can be treated
 as a Parker spiral field
\begin{equation}
 \vec{B}\sw(\vec{r}) = (\left.B_{\rm sw}\right|_{1\,\si{AU}}) \, \frac{r_{0}^{2}}{r^{2}}
 \left( \vec{e}_r + \frac{r \, \Omega\sw}{v\sw} \, \sin\vartheta \vec{e}_{\varphi}\right)
\end{equation}
into the inner computational volume. At the outer boundary, the homogeneous
ISM field is applied, and both fields meet in an intermediate transition region
of finite width, where a spatially weighted average of the respective vector
potentials establishes a smooth and divergence-free transition.

%H
%
\subsubsection{The plasma $\beta$}

The interstellar magnetic field strength of $10\,\si{\micro G}$ was chosen to be
distinctly above the average Galactic field of about $3\,\si{\micro G}$
\citep{Planck-Collaboration-2016}.
But even this rather high field strength
only causes a small asymmetry compared to other astrospheres, like
the highly asymmetric heliosphere. The reason is that unlike in thermal plasmas,
where the so-called thermal plasma beta
\begin{equation}
  \label{eq:1}
  \beta_{\rm therm} = \frac{8 \pi P}{B^{2}}
\end{equation}
determines the dynamics, for super-fast magnetosonic scenarios the thermal
pressure $P$ in Eq.~(\ref{eq:1}), in which the scalar pressure terms in
the momentum equation are compared, should be replaced by the ram pressure
$\rho v^2/2$. This results in
\begin{equation}
  \label{eq:2}
  \beta_{\rm ram} = \frac{8\pi}{B^{2}}\, \frac{\rho v^{2}}{2},
\end{equation}
where $\rho$ and $v$ are the
density and speed, respectively \citep[see][]{Scherer-etal-2016a}. The most general expression is therefore
\begin{equation}
  \label{eq:2a}
  \beta = \frac{8\pi \left(\dfrac{\rho v^{2}}{2}+P\right)}{B^{2}} .
\end{equation}
In the subsonic case, the thermal pressure dominates the numerator,
while in the supersonic case the ram pressure is higher. Thus,
Eqs.~(\ref{eq:1}) and (\ref{eq:2}) are upper and lower limits of
Eq.~(\ref{eq:2a}). It is also instructive to note that Eq.~(\ref{eq:2a})
may be written as
\begin{equation}
  \beta = \dfrac{v^{2} + (2/\gamma) \, v^{2}_{\rm s}}{v_{\rm A}^{2}} =
  \left(1 +\frac{2}{\gamma M\rms^{2}} \right) M\rmA^{2} ,
\end{equation}
where $v\rms$ and $v\rmA$ are the thermal sound speed and the Alfv\'en
speed, respectively, while $M\rms$ and $M\rmA$ denote the corresponding Mach
numbers.

This explains why the
astrosphere around \LC is more or less axially symmetric with respect to the
inflow axis. As shown by \citet{Scherer-etal-2016a}, the interstellar
$\beta_{\rm ram}$ is equal to 140, while
for the heliosphere $\beta_{\rm ram}= 0.8$. Therefore, the magnetic field
is dynamically important for the heliosphere and produces a strong asymmetry
\citep[e.g.][]{Scherer-etal-2016a, Pogorelov-etal-2017}.
Consequently, the large-scale structure of astrospheres around runaway stars
with high velocities relative to the ISM will be almost symmetric
like in HD scenarios, at least in the upwind direction.

\begin{figure}
  \includegraphics[width=\columnwidth]{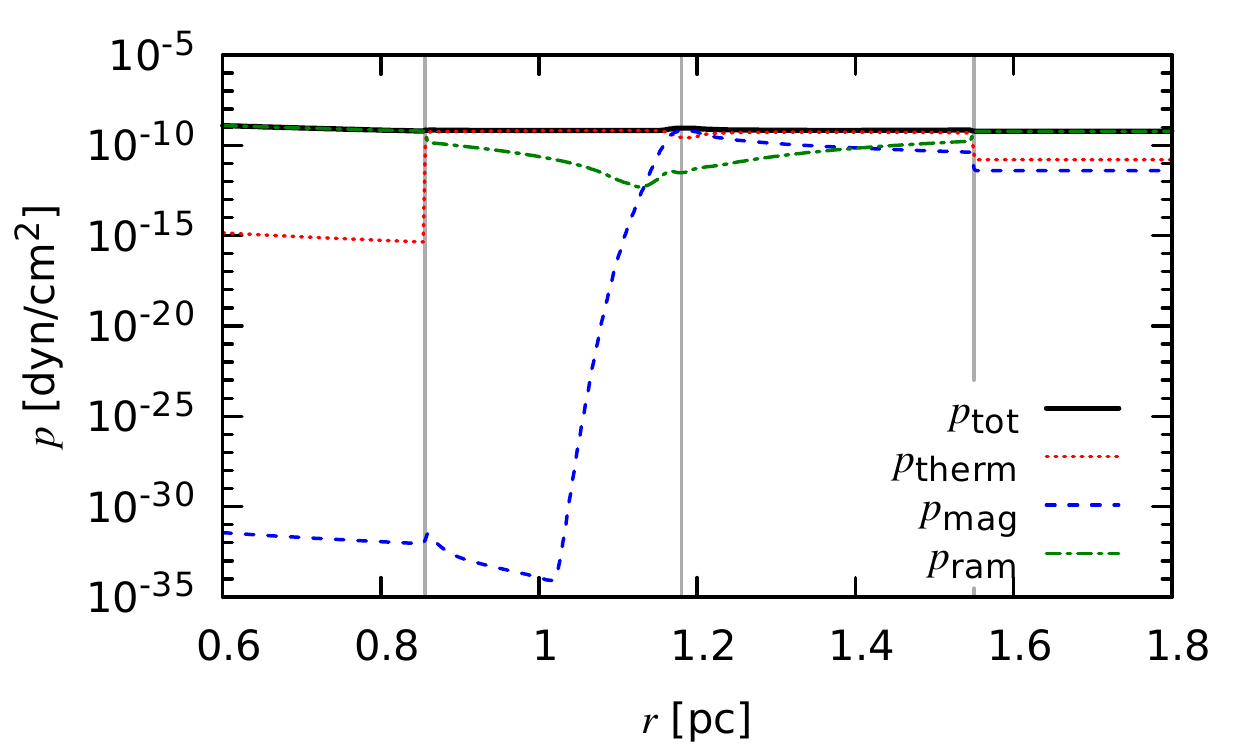}%Fig3p-crop.pdf}  
  \caption{ \label{fig:2}
    The profiles for \LC of total pressure (black solid line),
    thermal pressure (red dotted line),
    magnetic field pressure (blue dashed line), and
    ram pressure (green dash-dotted line)
    along the ``inflow axis''.  The vertical grey lines mark the positions
    of the TS, the AP, and the BS (from left to right).}
\end{figure}

For a more detailed description, Fig.~\ref{fig:2} compares the different pressure terms. As can be seen, the
total pressure is dominated by the ram pressure inside the TS and jumps by
a factor of four at the TS, which is in agreement with the Rankine-Hugoniot relations.
Moreover, it reaches very low values at the stagnation point on the AP. (It should
vanish on both sides of the astropause because the normal velocity is zero;
this, however, can not be realized due to numerical limitations.) Beyond the TS, over the AP towards
the BS, the thermal pressure dominates, with a small contribution of the
magnetic field pressure at the AP.

For the stellar magnetic field, which is assumed to be a Parker-like
spiral field, the magnetic pressure decays in the
ecliptic as $r^{-2}$ but over the poles as $r^{-4}$, and we find the
thermal plasma $\beta$ to scale as
\begin{equation}
  \beta_{\mathrm{therm}} \propto
  \begin{cases} r^{-4/3} & \mathrm{in\ the \ ecliptic}\\
    r^{+2/3} & \mathrm{over\ the\ poles}
    \end{cases}
\end{equation}
because the pressure decay is proportional to
$r^{-10/3}$. Thus, for subsonic conditions, the magnetic field
pressure in the ecliptic is larger than the thermal pressure, and vice
versa over the poles. However, in the supersonic case, we find
\begin{equation}
  \beta_{\mathrm{ram}} \propto
  \begin{cases} r^{0} & \mathrm{in\ the \ ecliptic}\\
    r^{+2} & \mathrm{over\ the\ poles}
    \end{cases}
\end{equation}
since mass continuity implies $\rho \propto r^{-2}$ for constant speed $v\sw$.

\subsubsection{Flow properties}
\begin{figure}
\includegraphics[width=0.495\textwidth]{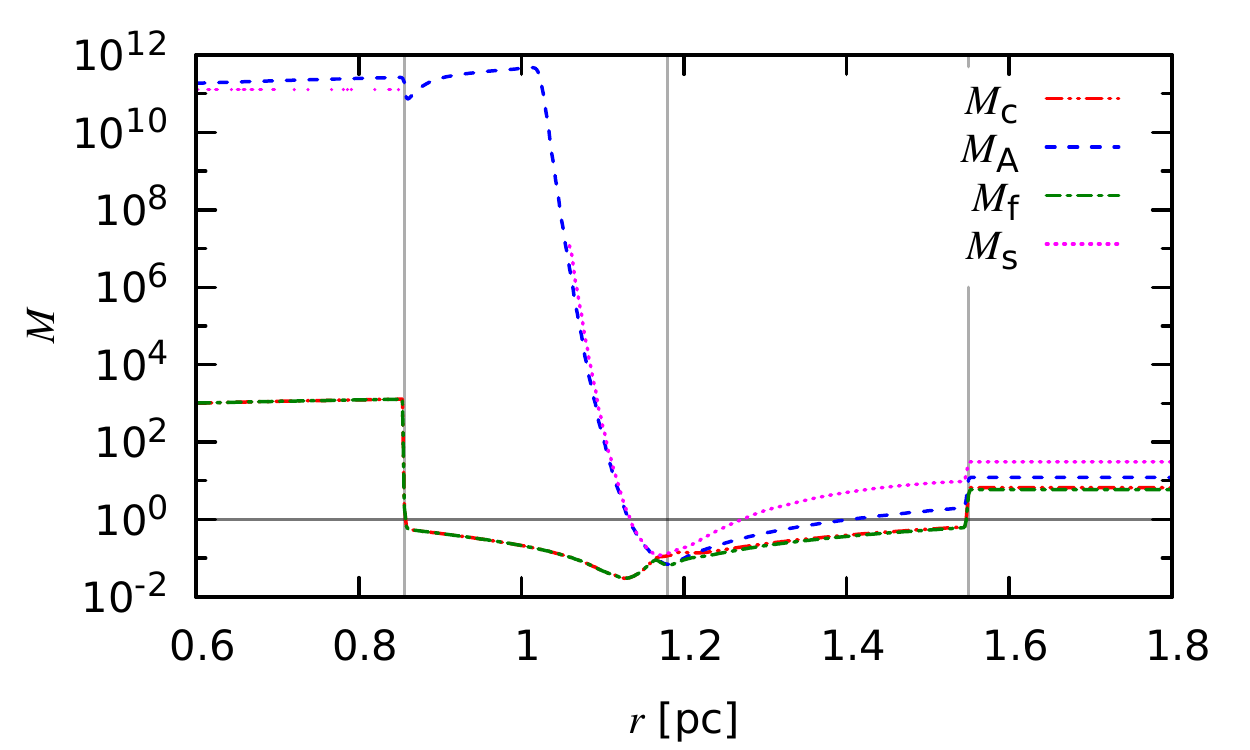} %cvall_new.pdf}
\includegraphics[width=0.495\textwidth]{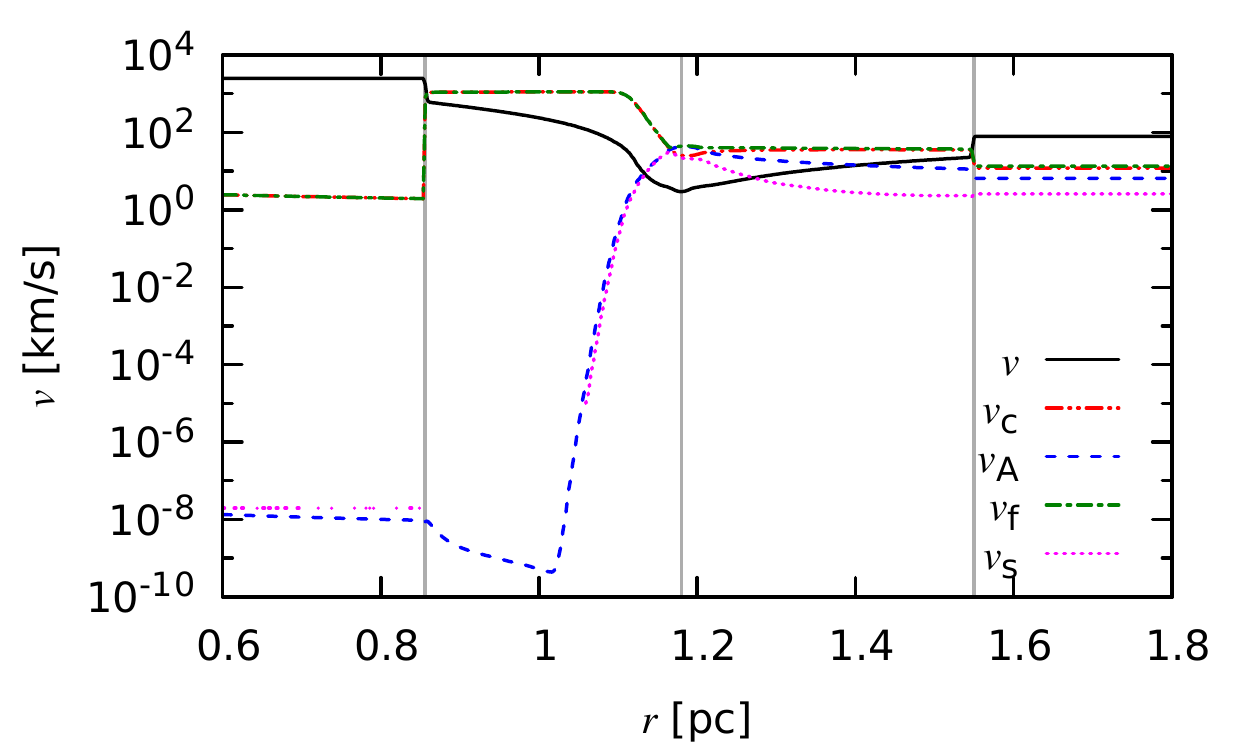}%3M-crop.pdf}
\caption{Profiles for \LC of characteristic speeds together with the actual
  flow speed $v$ along the inflow axis (left panel)
  and the corresponding Mach numbers (right panel). The vertical grey lines are from left to right the TS, AP and BS.}
\label{fig:3}
\end{figure}

To discuss the flow properties, it is convenient to consider the Mach numbers
for the different wave speeds, which are the sonic Mach number $M\rmc$,
the Alfv\'{e}n Mach number $M\rmA$, the fast magnetosonic one $M\rmf$, and
the slow magnetosonic Mach number $M\rms$
\citep[e.g.][]{Goedbloed-etal-2010}.
The corresponding speeds are denoted by $v\rmc$, $v\rmA$, $v\rmf$, and $v\rms$.
The profiles for these quantities, which have been extracted from the
simulations, are presented in Fig.~\ref{fig:3}. Note that the
speeds and Mach numbers are given in the observer frame. To calculate
the Rankine-Hugoniot jump conditions, these parameters have to be
transformed into the shock rest frame (see below). In Appendix~A, we present the Rankine-Hugoniot relations for the HD and MHD scenarios and compare these analytic results with those from our \LC model.
\begin{figure*}
\includegraphics[width=0.495\textwidth]{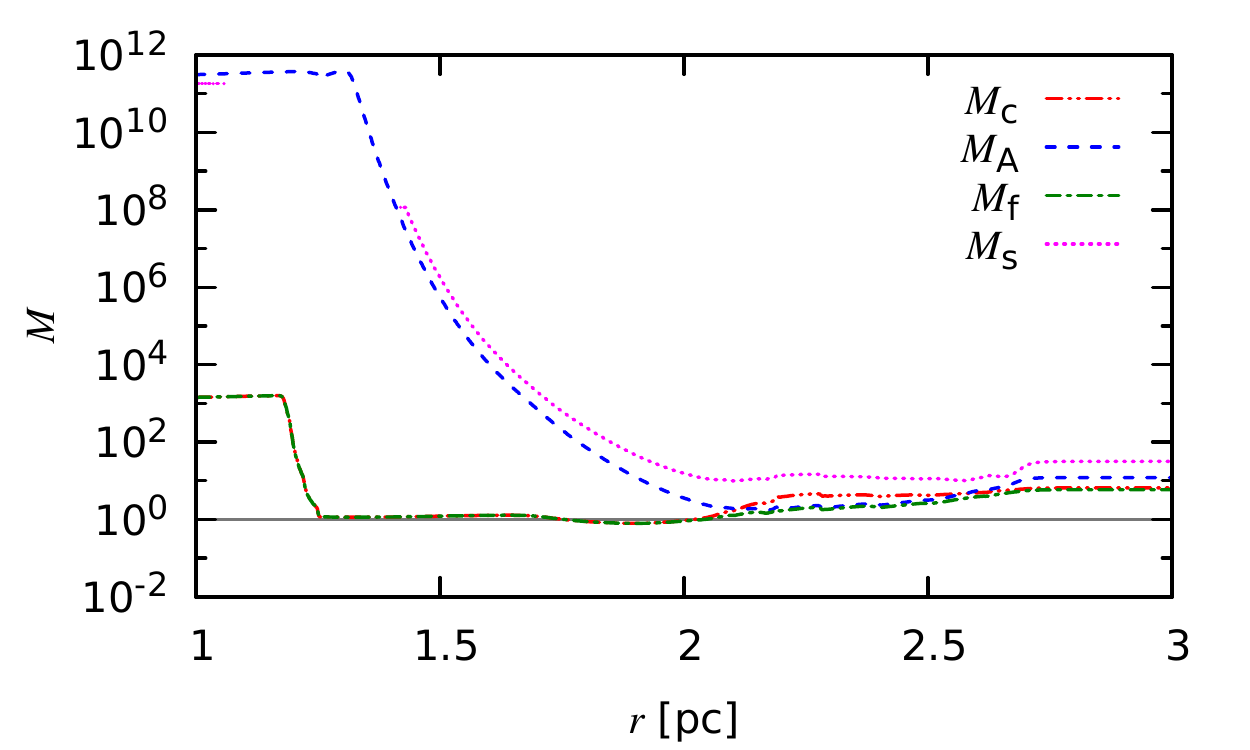}%cvall_new.pdf}
\includegraphics[width=0.495\textwidth]{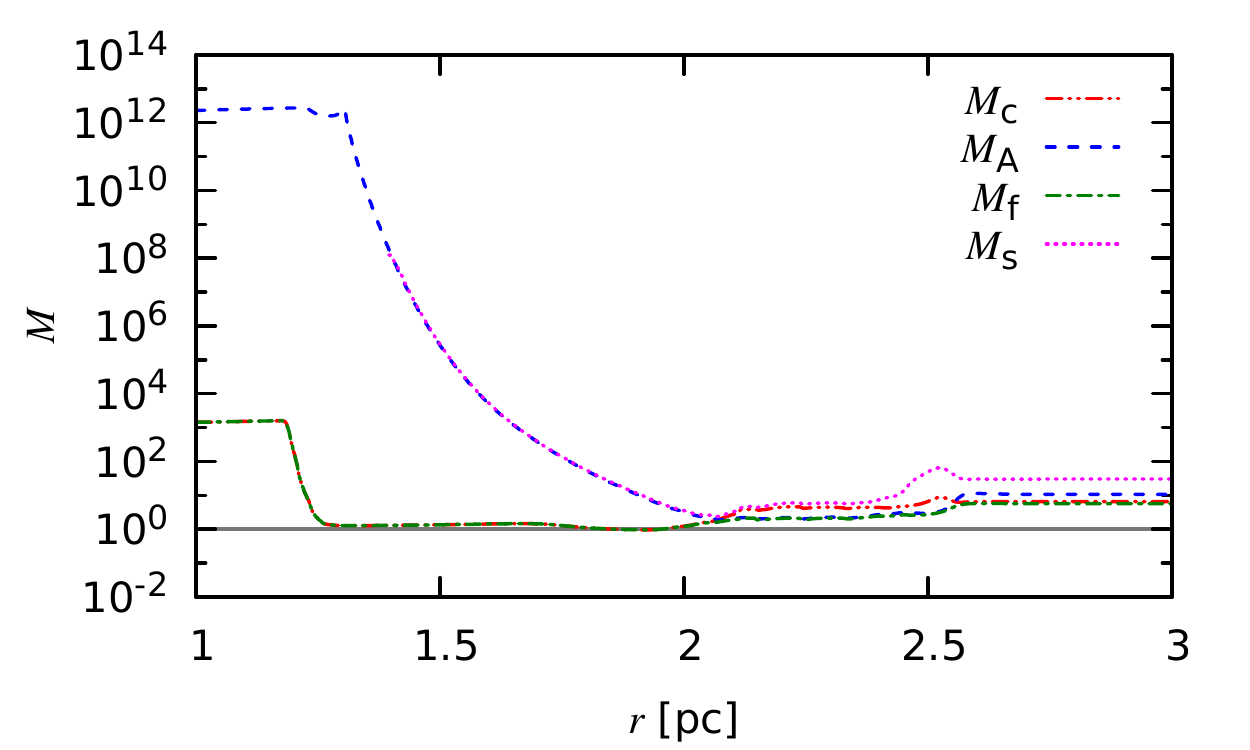}\\ 
\includegraphics[width=0.495\textwidth]{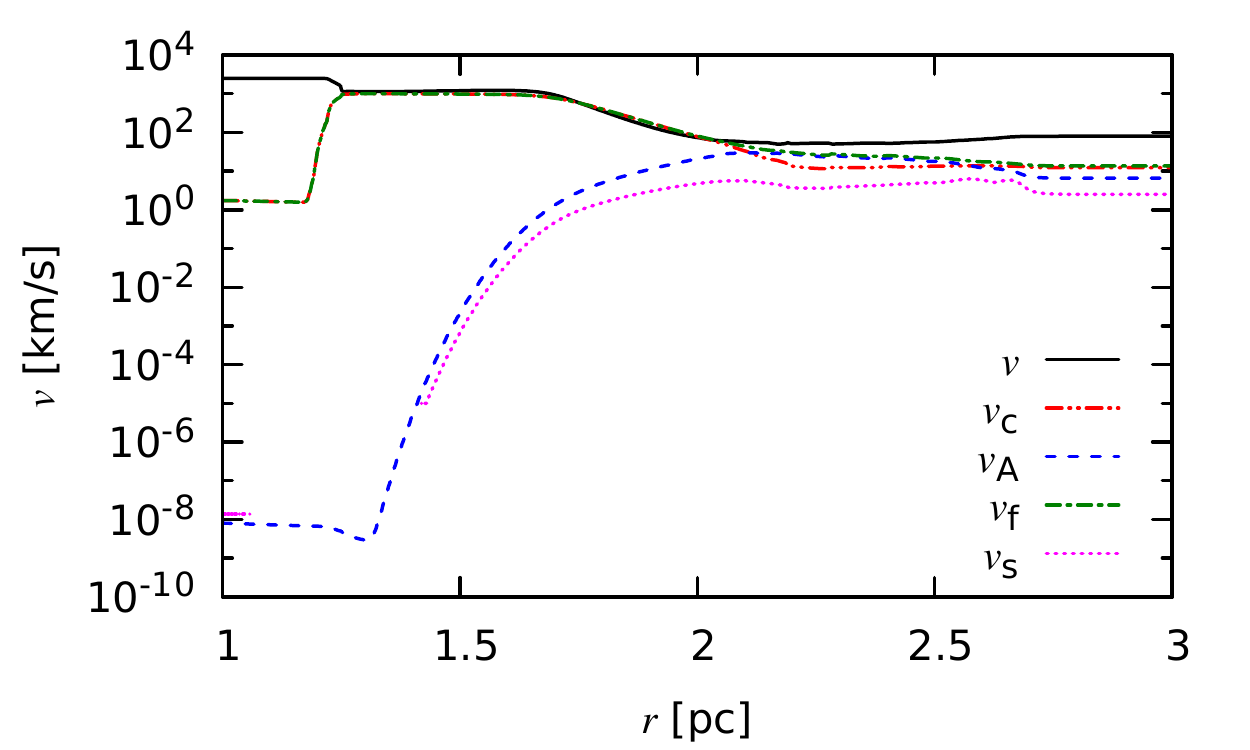}%3M-crop.pdf}
\includegraphics[width=0.495\textwidth]{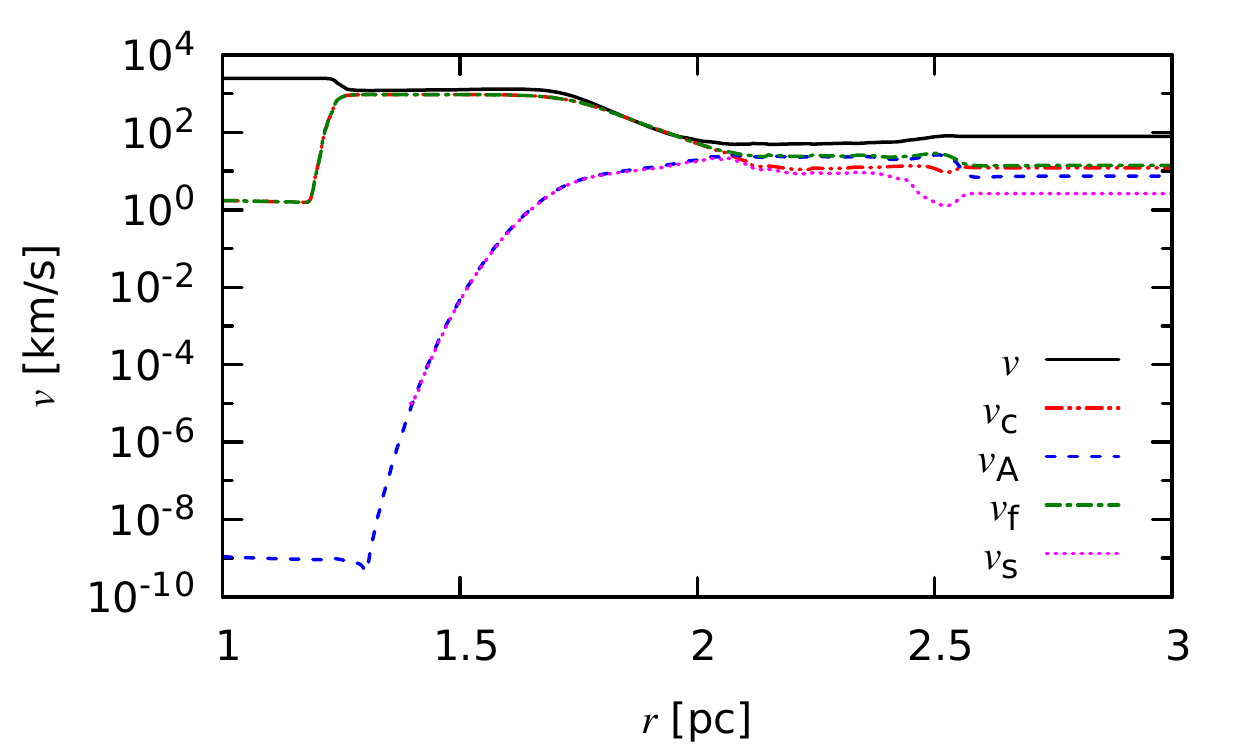}%3M-crop.pdf}
\caption{Similar to Fig.~\ref{fig:3}: the left panel is in the inflow
  plane, but with an angle of $90\si{\degree}$ with respect to the inflow
  direction, and the right panels show the parameters over the pole. }
\label{fig:3c}
\end{figure*}
As can be seen from Fig.~\ref{fig:3}, the flow speed shows the expected
behavior: It almost vanishes at the AP and jumps by a factor of four at the TS and BS.
The sound speed $v\rmc$ is lower than $v$ in the
 supersonic regime, inside the TS and outside the BS, and higher inside. The Alfv\'{e}n speed is constant inside the TS because the
spiral magnetic field in the ecliptic (in the model the plane containing the
stagnation line) drops as $r^{-1}$ while the density drops as $r^{-2}$ and thus the Alfv\'{e}n speed is constant along the stagnation line
up to the TS. Beyond the TS, it jumps slightly by a factor of two,
because $B$ jumps roughly by a factor of four and $\sqrt{\rho}$ by a
factor of two. The Alfv\'en speed then remains mainly constant and starts to increase towards the
AP, where it reaches its maximum value and decreases again towards the BS, where
another jump by a factor of two occurs.

The fast magnetosonic speed $v\rmf$ is mostly dominated by the sound speed, except
in the outer astrosheath, where the sound speed and Alfv\'{e}n speed are of the
same order. $v\rms$ is almost zero inside the TS, jumps at the TS, and drops to zero
in a region of the inner astrosheath. It reaches its maximum at the AP and
then decreases until it jumps at the BS to the unperturbed interstellar value.

\begin{figure*}
\includegraphics[width=0.49\textwidth]{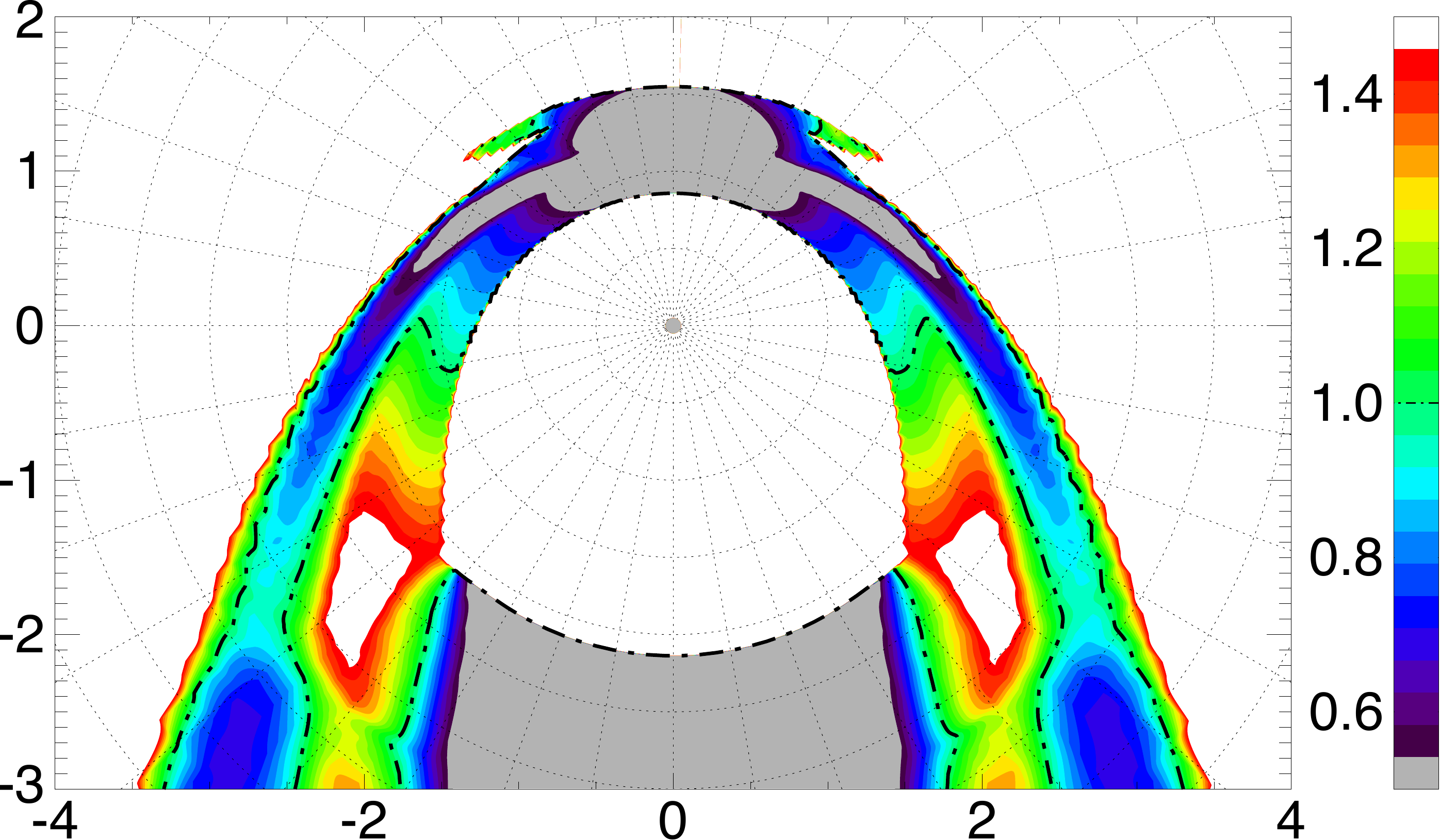}
\includegraphics[width=0.49\textwidth]{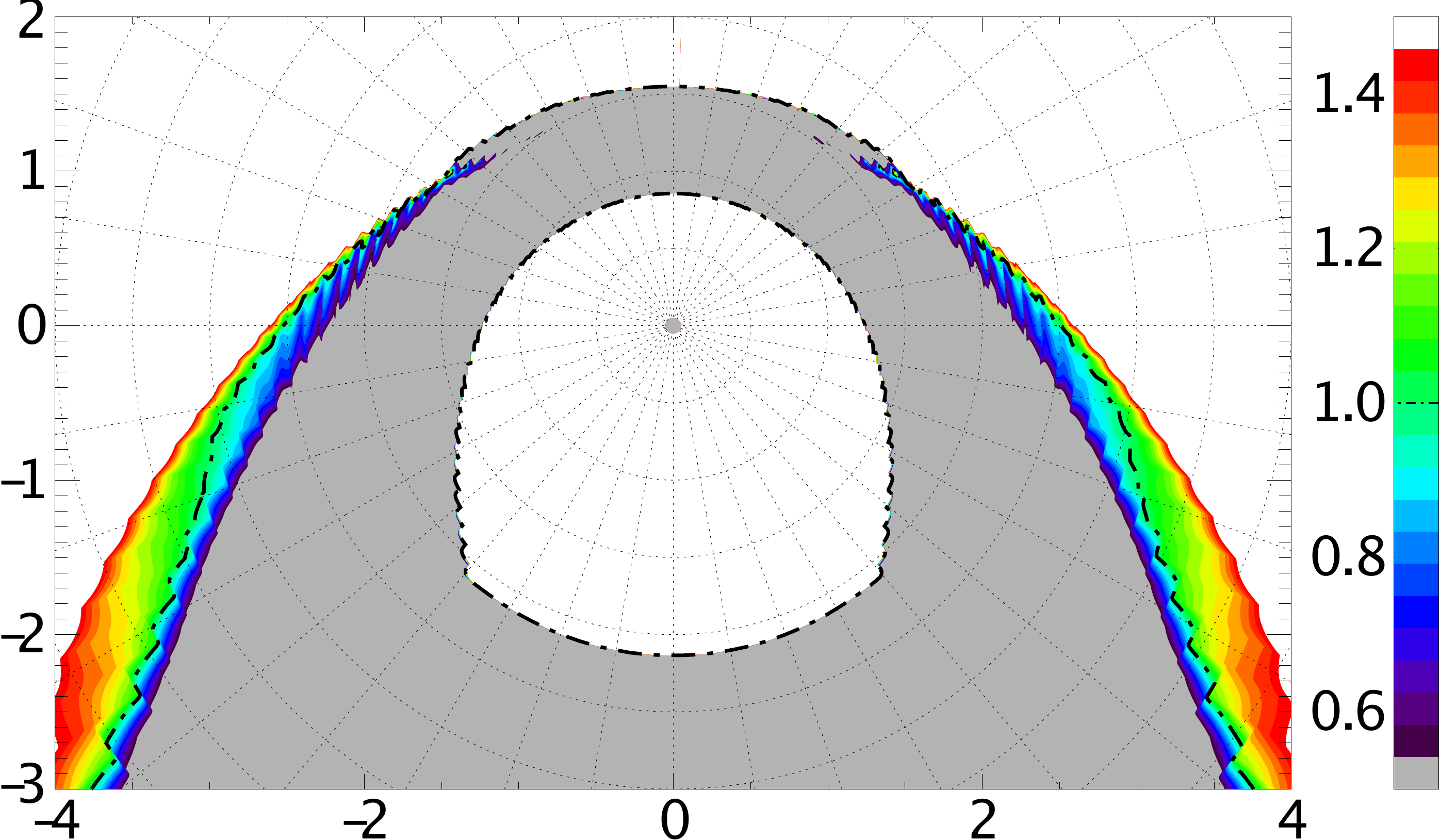} \\
\includegraphics[width=0.49\textwidth]{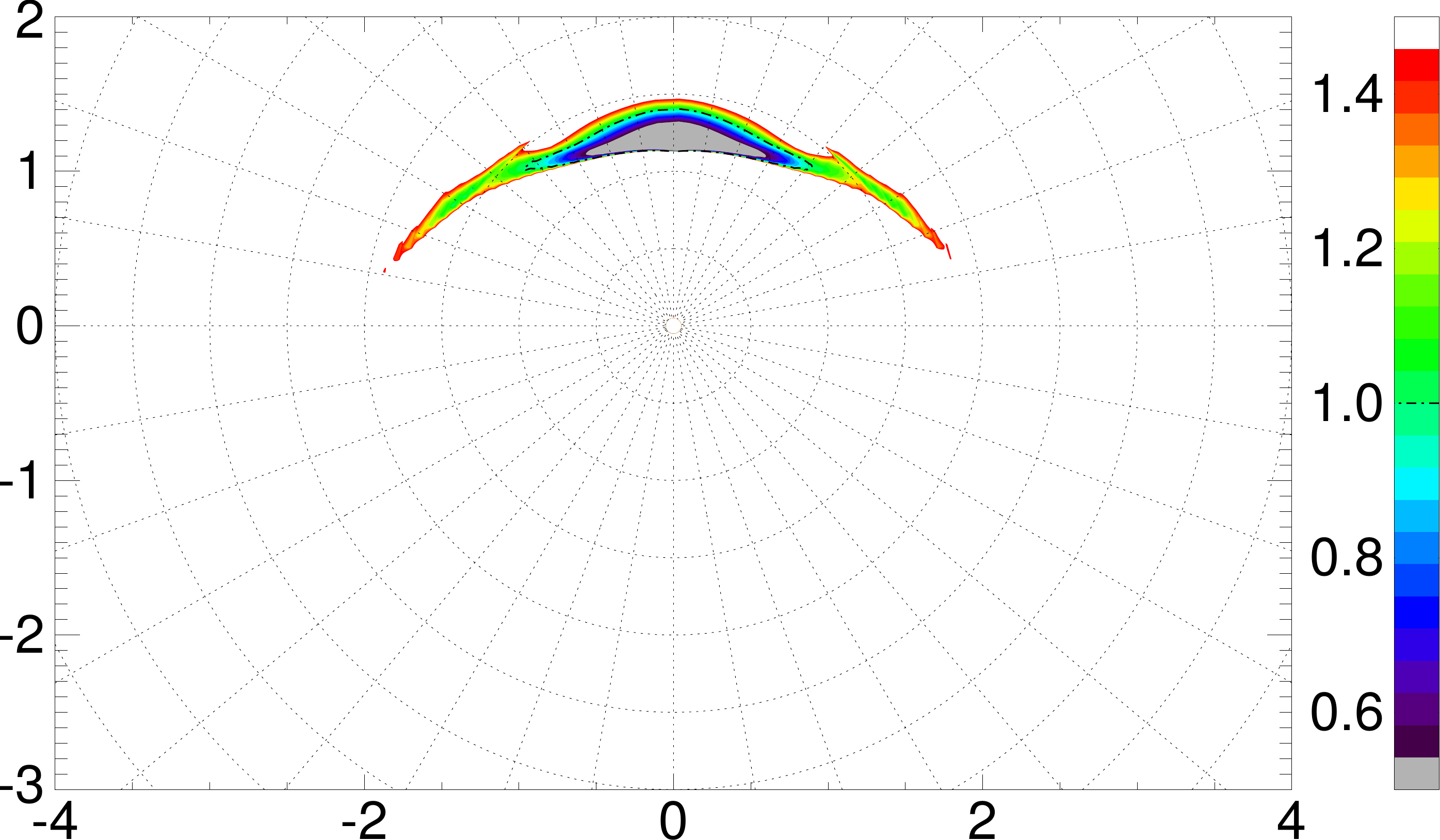}
\includegraphics[width=0.49\textwidth]{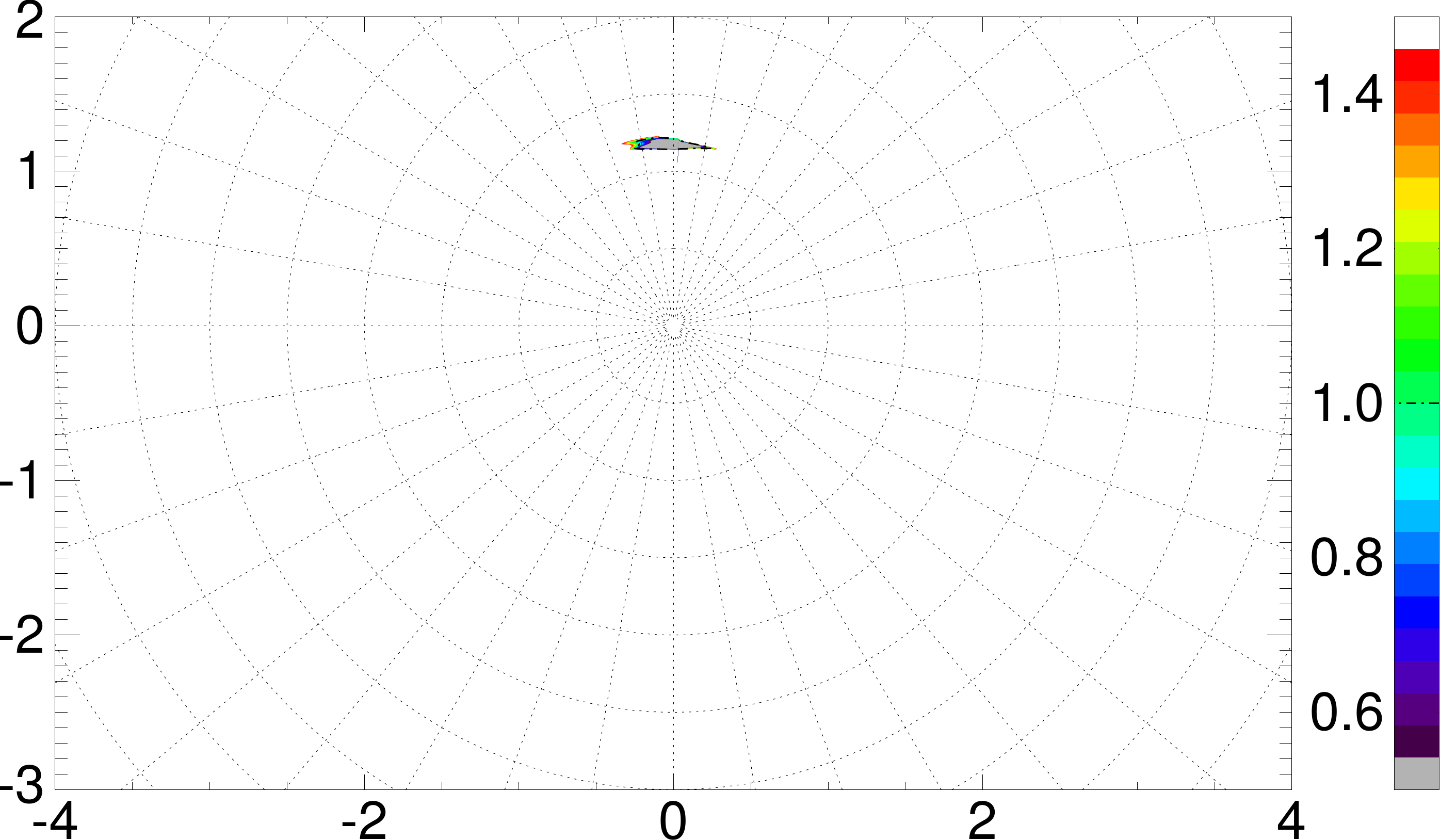}
  \caption{{For the \LC model, }the four Mach numbers $M\rmc$, (upper left panel), $M\rmf$
    (upper right), $M\rmA$ (lower left), and $M\rms$ (lower right) in
    the range between 0.5 and 1.5. The dashed black lines mark the transitional
    value $M_{i}=1, i\in\{{\rm c}, {\rm f}, {\rm A}, {\rm s} \}$.
    In the white areas the Mach numbers exceed 1.5, while in the grey areas they are smaller than 0.5. Length scale in pc.}
  \label{fig:4}
\end{figure*}

As can be seen from the left panel of Fig.~\ref{fig:3}, $M\rmf$
and $M\rmc$
are almost identical. They are higher than one inside the
astrosphere and in the unperturbed ISM, and smaller than one
elsewhere. Both $M\rmA$
and $M\rms$
are always greater than one, except for a narrow region at the AP.
Close to the AP, there is a region in which $M\rms$
drops below one.  Thus, a transition from a super- to a
sub-slow magnetosonic flow in the flanks can be expected. Thus, transitions can indeed be seen in
the panels of Fig.~\ref{fig:3c} and Fig.~\ref{fig:4}, in which these regions are plotted in the ecliptic plane.
The upper and lower panels of Fig.~\ref{fig:3c} show the quantities $M$ and $v$, respectively. While the left panels display the results in the flank
  (that is the direction $90\si{\degree}$
  off the inflow direction), the right panels show the results
  above the north pole (i.e., the direction $90\si{\degree}$
  above the ecliptic). As can be seen in Figs.~\ref{fig:3} and~\ref{fig:3c},
  there are subtle differences in the Mach numbers as well as in the characteristic speeds.  The transition at the TS in
  the flank and over the pole is more or less the same as that along
  the inflow axis (see Fig.~\ref{fig:3}), while the transition at the BS 
  is much weaker.  The latter is caused by the fact that in these regions the shock is more oblique and thus, the Mach numbers decrease, as do the compression ratios. The latter can be found by solving the Rankine-Hugoniot equations, or directly from the model.

While detailed knowledge of these structures is quite helpful for
the understanding of cosmic-ray transport models, a discussion of the
latter is beyond the scope of this work
\citep[but see][]{Scherer-etal-2015c, Pogorelov-etal-2017}.  The importance of the Mach numbers was already discussed in \citet{Webb-etal-1986},
  while in the recent literature
  \citep[][and references therein]{Pogorelov-etal-2017}, mainly numerical models were used to model the cosmic ray transport. The latter requires the knowledge of the diffusion tensor and the drift coefficients, both strongly depending on the fluctuations and the compression ratio (Mach numbers), as discussed in, for example, \citet{Schlickeiser-2002}. 
 The detection of non-thermal emission from a runaway O-star was
 reported by \citet{Benaglia-etal-2010}, see also \citet{delValle-Pohl-2018}.

Fig.~\ref{fig:4} shows contours of the four relevant Mach numbers
$M_{i}$, $i\in\{{\rm c}, {\rm A}, {\rm f}, {\rm s} \}$ in the ecliptic in
an interval centered on the respective transitions $M_{i}=1$.
The latter is represented by a black dashed line, which can easily be
seen for $M\rmc$ (upper left panel of Fig.~\ref{fig:4}) and hardly
in the case of $M\rmf$ (upper right panel of Fig.~\ref{fig:4}). The
sonic Mach number $M\rmc$ behaves similarly as in the HD case: There is
an additional tangential discontinuity emanating from the triple
point in the tail direction. The triple point may be identified as the sharp
edge in the tail. There is another sonic line closer to the inflow direction,
which also extends into the tail direction. This is different from the
stationary HD case where it matches the AP and then extends towards the BS.
The third line with $M\rmc=1$ starts in the flanks and lies close to the BS.
There is also a $M\rmc<1$ region near the nose.
The oblique shocks in the flank region are usually weak shocks, i.e., \
$M\rmc >1$ behind the shock. Although the plasma in
the outer astrosheath is strongly cooled, the temperature is nevertheless
still high enough to enforce $M\rmc<1$ everywhere except in the flank region
toward the tail.

The Alfv\'{e}n Mach number $M\rmA$
is always much higher than expected in an area around the nose
direction (see lower right panel of Fig.~\ref{fig:4}.% and the enlarged view in the lower left panel of Fig.~\ref{fig:5}). 
As can be seen, $M\rmA$ maintains higher values shortly after the BS but then drops below one before increasing again towards the AP.
In the region where $M\rmA<1$, the flow becomes subslow
magnetosonic, and thus the magnetic field decreases to lower values
(see below).
In the upper right panel of Fig.~\ref{fig:4}), it is evident that the
the transition region is almost infinitesimally small in the nose region.
Interestingly, there is also a small area in the nose region in which
$M\rms<1$ (lower right panel).
% in Figs.~\ref{fig:4} and \ref{fig:5}).  
%For convenience, the temperature and density in the enlarged region
%around the nose are shown in panels e and f of Fig.~\ref{fig:5}.

The discussion given above shows that \LC-like astrospheres with magnetic fields have a complicated structure, indicating that studies of
cosmic-ray propagation within and through such a structure require further investigations \citep{Scherer-etal-2015a,Scherer-etal-2016c}. 

As can be seen from Fig.~\ref{fig:2} and Table~\ref{tab:ts}, the magnetic pressure does not play any role for the dynamics, and could easily be set to zero. We nevertheless retained its non-zero value because it does not effect the overall computation time.

\subsection{Heliosphere-like astrospheres and wind bubbles}

One of the main differences between the astrospheres of hot and cool
stars is that hot stars can ionize their surroundings, while cool
stars are not able to do so. Therefore, in cool star scenarios like,
for example, our heliosphere, a neutral component and its interaction
with the ions must be taken into account. Besides, their astrospheres
are smaller because the ram pressure the stellar winds of cool stars
is usually much lower than that of hot stars. Thus, the cooling by
photons does not play such a crucial role like, for example, in the
large astrospheres around O-stars.  To compare such a scenario with
the O-star models, we neglected the influences of the neutrals, and
call it heliosphere-like scenarios.  The BS of a hot star is not
affect by neutrals, because the astrophere (except of the tail) is
inside the Str\"omgren sphere, and thus all particles are ionized.
Stellar wind bubbles, on the other hand, are a special subclass of
astrospheres. Here, no relative motion between the star and the
ambient medium exists. Thus, the bow shock is moving outward (away
from the star) until its speed reaches the local sound speed
(magnetosonic speed). After becoming a bow wave
\citep{Pogorelov-etal-2017}, it vanishes completely, corresponding to
a subsonic inflow for astrospheres. These scenarios are also
numerically harder to handle because disturbances can reach the outer
boundary, which makes it hard to define the values at "infinity."

The cooling depends strongly on the density of the plasma, thus in other scenarios, like in dust-driven winds \citep{Lamers-Cassinelli-1999} the cooling can take place at the TS  as well as on the BS. For line-driven or solar-like winds the number density at the TS is usually so low that cooling does not take place.

\subsection{Comparison of the different scenarios}
In the following, a heliosphere-like astrosphere, four \LC-like ones (with and
without cooling), and one wind bubble without relative motion with respect to the ISM (see Table~\ref{tab:c1}) will be studied in more detail,  \citep[see also][]{vanMarle-etal-2014}, where an HD and MHD simulations are discussed. There also the difference between the unstable "bow shocks" in HD and the "stable" ones  in MHD can be seen. As discussed above, this is caused by the notoriously unstable tangential discontinuities in HD, but the stable astropauses in MHD. 
\begin{figure}
  \includegraphics[width=0.45\textwidth]{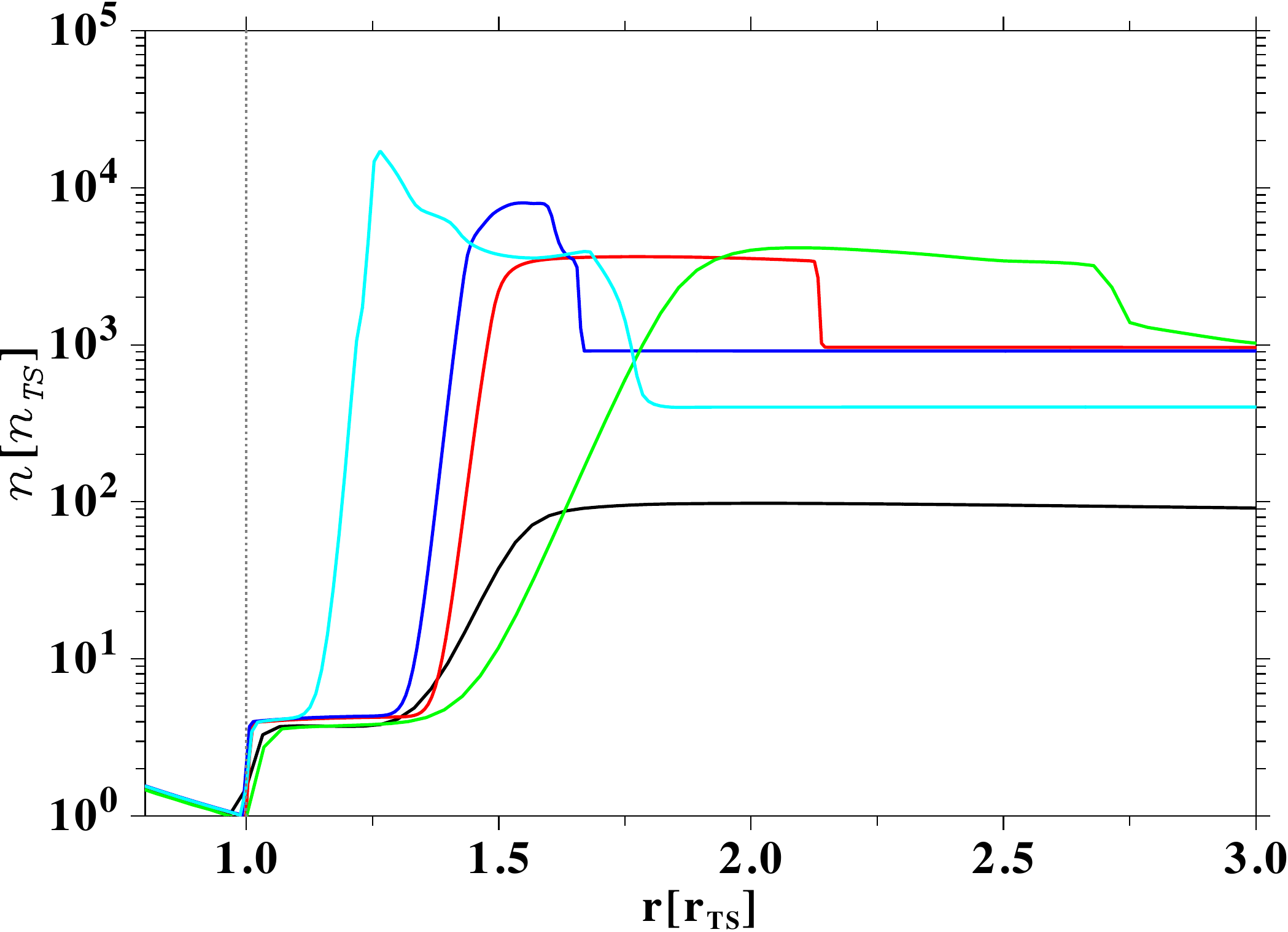}\\
  \includegraphics[width=0.45\textwidth]{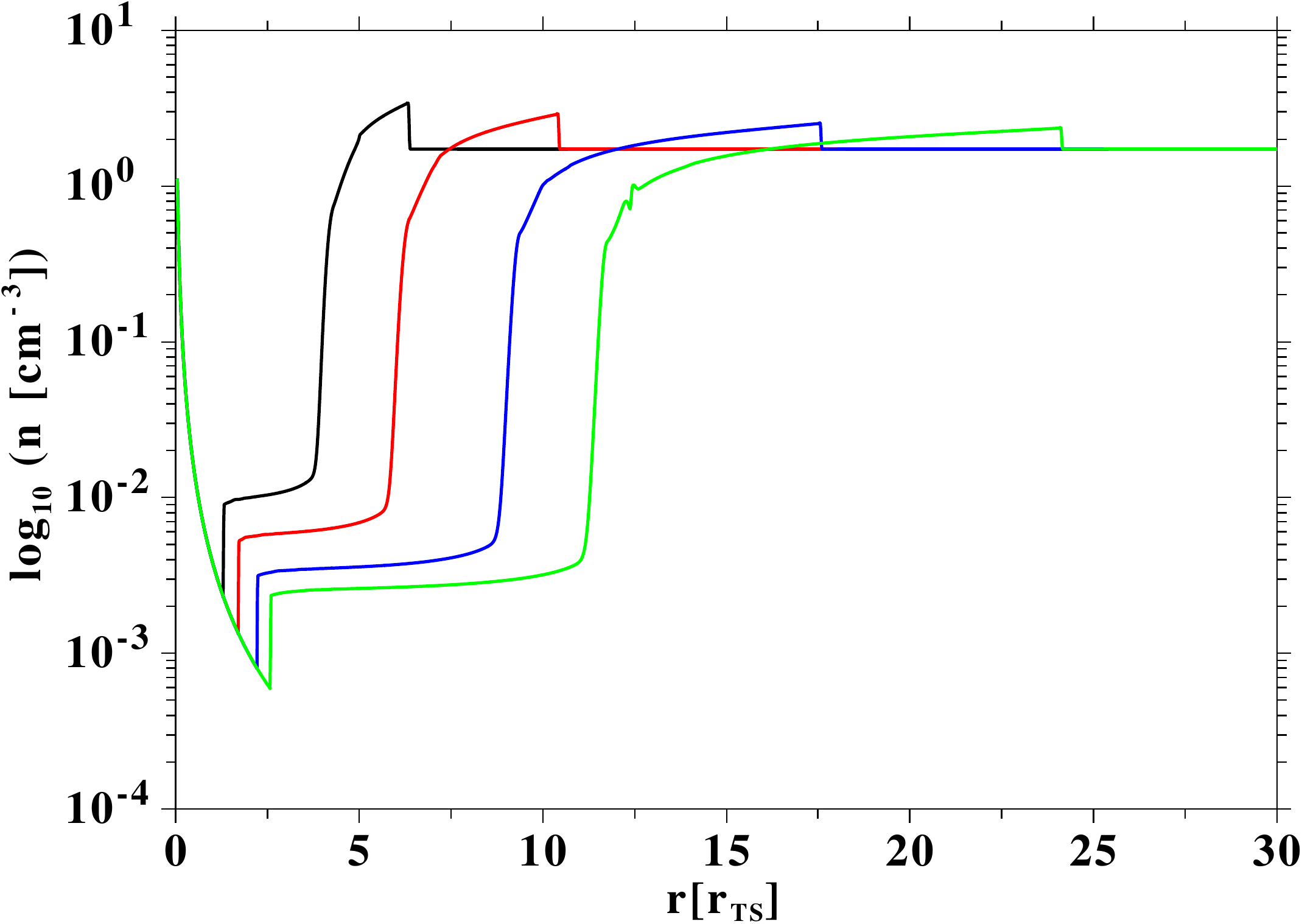}\\
  \includegraphics[width=0.46\textwidth]{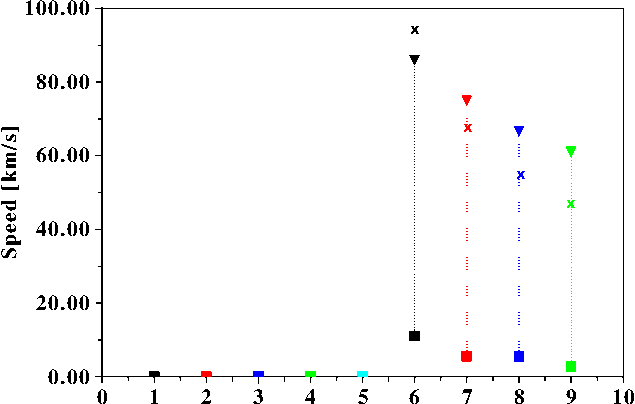}
  \caption{ Upper panel: number density as a function of distance,
    both normalized to values at the TS. The colored solid lines have
    the initial parameters for \LC.
    The black line indicates a heliosphere-like astrosphere (see
    Table~\ref{tab:c1}). The color coding is as follows.  Green: $B=0$
    without cooling, cyan: $B=0$
    with cooling, red: $B>0$
    without cooling, blue: $B>0$
    with cooling.  The TS position is indicated by the vertical dotted
    line.  The next sudden increase occurs at the AP, and the
    step-like decrease at larger distances is the BS. As can be read
    off, the TS is a strong shock with a compression ratio of 4.
    The middle panel shows the wind bubble model at times
    $t=50$\,kyr (black), $t=100$\,kyr (red),  $t=200$\,kyr
    (blue), and $t=300$\,kyr.
    The motions of the TS, AP and BS are nicely seen. Lower
    panel: Corresponding shock speeds at the TS (squares) and at the
    BS (triangles), using the same color coding as on the upper panel,
    but the bubble models are shown as dotted lines. The cross indicates the HD analytically determined BS speed for the energy-conserving snow-plow phase.}
  \label{fig:6}
\end{figure}
It is well known that the ideal MHD equations can be made dimensionless by
normalizing length $L$, mass density $\rho$, velocity $v$,
magnetic field amplitude $B$, and thermal pressure $P$ according to 
\begin{equation}
  \label{eq:n1}
  \begin{array}{rl@{\quad}rl@{\quad}rl}
    L =& l_{0} \, L^{\prime} = r\TS \, L^{\prime}, & \rho =& \rho_{0} \, \rho^{\prime}, & v =& v_{0} \, v^{\prime},  \\
    && B =& B_{0} \, B^{\prime}, & P=& \rho_{0} v_{0}^{2} \, p^{\prime} ,
  \end{array}
\end{equation}
using normalization constants (indicated by a lower index ``0''). Thereby, $v_{0}$ is most often chosen to be the Alfv\'{e}n speed
\mbox{$v_{{\rmA},0}=B_{0}/\sqrt{4\pi\rho_{0}}$} or, if $B=0$, as the sound
speed of ideal gases. The time scale then follows from the length scale as
\mbox{$t_{0}=l_{0}/v_{0}=r\TS / v_{0}$}.
Thus, all astrospheric models based on ideal (M)HD are self-similar and
scale-invariant if cooling is neglected. This scale-invariance is
violated by the inclusion of cooling or neutrals that interact with the plasma via charge exchange. Other effects like viscosity or resistivity that usually are not considered in astrosphere models, will also break the scale-invariance, i.e., they require the introduction of dimensionless
parameters such as the magnetic or viscous Reynolds numbers.

Nevertheless, we will use the normalization constants
\mbox{$\rho_{0}=\rho(r\TS)$}, \mbox{$B_{0}=10\,\si{\micro G}$}, and \mbox{$l_{0}=r\TS$} to show the normalized number density along the inflow axis (see upper panel of Fig.~\ref{fig:6}). One should keep in mind that the actual
value of $r\TS$ is different for all models shown here. From the upper panel of
Fig.~\ref{fig:6} it shows that astrospheres without magnetic fields
and cooling (green line) extend deeply into the ISM, while those without magnetic fields
but with cooling (cyan line) show a stronger shrinking of both the inner and outer
AP, and also features a high-density peak in front of the AP. Note that the
height of this density peak depends on the applied cooling function
(not shown here). In the next two models, the magnetic field is switched on,
first without cooling (red line) and then with cooling (blue line). The
model without cooling also extends deeply into the ISM,
but the number density in the outer astrosheath remains nearly the
same, and the BS moves further inwards due to the non-zero magnetic
field pressure. For the case with cooling,
the BS also moves inward, but the density peak is now lower
compared to the non-magnetic case. The magnetic
fields also stabilize the position of the AP: both APs are of similar
extent in this case, while in the case without a magnetic field, the
presence of cooling clearly separates the AP extensions.

Moreover, the interstellar density in the models without magnetic
fields and cooling seems to be lower than that of the
other \LC models. This is merely an artifact of the
normalization to a different $r\TS$, which can more clearly be seen in the
wind bubble model (middle panel), in which the bubble, still expanding into the ISM,
causes the TS to move to larger distances. However, since the density then
decreases with $r^{-2}$, the normalized values of the ISM density change.

In the wind bubble model, one can clearly see that the relative density
increases with time for the same reason: the TS moves outward, and the stellar wind density decreases further towards the TS while the absolute ISM density remains constant. Thus, the normalized density has to increase. It is also shown that the
BS moves outward relative to the TS, and so does the
AP. Towards the end of the calculation, the speed of the BS
becomes lower (see lower panel).

 We note that if the normalized values 
for different scenarios are the same or similar, the above solutions
can be used simply by converting the TS radius in physical units. This does not hold if different normalized initial values are used.
Moreover, the expansion velocity $v_x$ obtained by HD conservation laws differs remarkably from those shown in Fig.~\ref{fig:6}. For the energy-conserving snow-plow phase we have:
\begin{align}
    v_x = 16.8 \left(\frac{L_{36}}{\mu \, n_{\rm ISM}}\right)^{1/5}t^{-2/5}_6\quad \mathrm{[kms/s]}
\end{align}
where $L_{36}=10^{-32}\dot{M}v_{\infty}^2$ and $t_6 = t \times 10^{-6}$\,yrs
\citep[see for example][]{Lamers-Cassinelli-1999}. For the wind bubble parameters from Table~\ref{tab:c1}, this gives $n_{\rm ism}=1.7$\,part./cm$^-3$ and a mass-loss rate $\dot{M} =  6.8\cdot 10^{-6} M_{\odot}$\,/yr for \LC, where $M_{\odot}$ is the solar mass.
Inserting the time steps as used in Fig.~\ref{fig:6}, we get different expansion speeds for the BS, i.e.\ for $t=50, 100, 200$, and 300 kyr we get $v_x = 95, 72, 55, 46$\,km/s. These speeds differ from those obtained by our model because of the additional magnetic field, which also causes a non-spherical shape of the corresponding astrosphere \citep[see][]{vanMarle-etal-2015}. These velocities are indicated in Fig.~\ref{fig:6} by crosses.

The shocked stellar wind density in the inner astrosheath, defined as
the region between the TS and the AP, is constant. This is true for all models,
because in the inner astrosheath the fluid is nearly incompressible,
which means that also the pressure is more or less constant, except
for a small influence from the flow speed according to the Bernoulli
equation in the HD case. 

In the MHD case, this is not a priori clear, but the simulation apparently leads to only small corrections because of the
weak magnetic field in this region. The reason is that the
magnetic field decreases in the ecliptic as $r^{-1}$ but jumps at the TS
only by a factor of four, while the thermal pressure jumps by orders of
magnitudes and is much higher in the inner astrosheath compared to both the
ram pressure and the magnetic field pressure. Thus, the behavior is more
HD-like in that region. Finally, we note that the heliosphere-like astrosphere (black line) has a subsonic
inflow speed, and thus forms no bow shock.
It can be seen in all models that the density jump (or
compression ratio $s$) reaches its maximum value of $s=4$ both at the TS
and the BS.

The TS and the BS are both fast magnetosonic shocks, while in the region in front of the astropause, there is a region bounded by a slow magnetosonic shock. This region, in particular, deserves more attention since it is equivalent to a huge magnetic island. Similar structures can be found in the heliosphere, see \citet{Opher-etal-2012}. Studies at a much higher resolution are currently under development. However, see the
  discussion by \citet{Ha-etal-2018} for the difficulties to
  accelerate protons at quasi-parallel slow shocks. 
\section{Shock speeds}
The shock speed
\begin{equation}
  \label{eq:rh2}
  u =
  \frac{\rho_{1} \, v_{1} - \rho_{2} \, v_{2}}{\rho_{1}-\rho_{2}}
\end{equation}
can be easily calculated from the Rankine-Hugoniot equation derived from the continuity-, momentum-, and energy equation, where indices 1 and 2 refer to the regions seen in the stellar rest frame, opposite to those described in Fig.~\ref{fig:6}, respectively. As can be seen, a
negative speed means that the shock speed is directed towards the
observer, whereas the sum of the shock speed and the stellar wind speed is
directed outward. For some models, for example, those with high speeds
($v\sw \approx 1000$~km/s)
and low densities at the TS ($n\sw \approx 10^{-3}$\,cm$^{-3}$),
the numerical evaluation of Eq.~(\ref{eq:rh2}) may cause problems.
It is then more advantageous to
calculate the shock speeds from consecutive time steps with sufficiently
large separation. The shock speeds for some models (calculated with the latter
method) are shown in the bottom panel of Fig.~\ref{fig:6}.
From Fig.~\ref{fig:6} it can be seen that the TS and BS speeds are
consistent with zero in the cases of astrospheres with relative
supersonic motion, which can be used to get a handle when the
numerical models reach stationary conditions.

In the case of the wind bubble models, it can nicely be seen that
the TS and the BS are moving outward at different speeds. While
the TS reaches a nearly constant speed between 50 and 200\,kyr, the BS only
reaches a stationary state after about 500\,kyr. The speed of the BS is
decreasing for all displayed times. From the middle panel of Fig.~\ref{fig:6},
it can also be seen that the compression ratio (shock strength) decreases
with increasing time. Thus, we can conclude that the BS will move
outwards until the compression ratio reaches unity, and the BS becomes a
sound wave.

Finally, the lower panel of Fig.~\ref{fig:6}, in addition, shows that if the
bubble model is still in evolution, its outward expansion slows down
and will eventually become a sound wave and the bow shock structure
will disappear. However, the time scales involved exceed 1\,Myr, which is close
to the lifetime of massive stars.  Thus, beyond that time, one can
expect other effects like, for example, a change in the stellar
wind speed to become important. The shocks in the upwind direction of the astrosphere
models are at rest, i.e., \ are stationary.
\section{Conclusions}

In this paper, the analytic 3D (M)HD structure of different astrospheres
were described for the first time, and regions with different
characteristic speeds have been identified. The knowledge about these
regions and their properties is important, for example, for the description
of turbulence and the propagation of cosmic rays in these cavities
\citep{Scherer-etal-2015a,Scherer-etal-2016c}.

\citet{Webb-etal-1986} already showed the importance of Mach numbers for cosmic ray acceleration in MHD models and \citet{Scherer-etal-2015a} used for simplicity the parameters derived from the heliosphere.  In this study, we, for the first time, have shown the Mach numbers to determine the shock structure or even its existence: If the fast magnetosonic Mach number of the ISM is lower than one, no bow shock will exist as discussed for the heliosphere \citep{Pogorelov-etal-2017b}. The Mach numbers also are used to determine the MHD shock type, especially for the existence of a fast or possible slow shock. However, to include the above-discussed Mach numbers to recalculate the diffusion tensor requires a reanalysis of the relevant turbulence parameters, see, for example, the recent work by \citet{Moloto-etal-2018}. 

Based on both theoretical arguments and simulations in full 3D, we demonstrated that the spatial shock structure of astrospheres such as the one around \LC is approximately axially symmetric. That implies that it would indeed be reasonable for follow-up studies to use high-resolution 2D settings instead, at least as long as their focus is not on issues related to, for example, magnetic structure and field lines which are indeed three dimensional, as can be seen in Fig.~\ref{fig:1}. 
  
We also determined the motion of the bow shock, which, for models
without relative motion between the star and the ISM, diminishes with
time. The BS continues to move away from the termination shock,
which is also not yet stationary. In the scenarios with no relative
motion, it was shown that the speed of both the TS and the
BS approach to constant but non-zero values on a timescale of about
$10^5$~years which is consistent with the recent literature \citep[for
example][]{vanMarle-etal-2014}. For the astrospheres with high relative
velocities with respect to the ISM, some of these speeds
are still high, indicating that the model has not yet reached stationary conditions. 

We have also shown that the \Ha glow mainly comes from a region where
the shocked interstellar density is highest, which occurs neither at
the BS nor at the astropause, but close to the latter, creating a hydrogen wall. The contribution from the shocked but
extremely hot stellar wind is negligible because  its number density
is too low.

\section*{Acknowledgements}
  KS, HF, and JK are grateful to the
  \textit{Deut\-sche For\-schungs\-ge\-mein\-schaft (DFG)}, funding the
  projects FI706/15-1 and SCHE334/9-2. DB and KW were supported
  by the DFG Research Unit FOR 1254. SESF acknowledges the partial financial support of the NRF under grants 93546 and 109253. KH acknowledges the International Space Science Institute and the supported International Team 464: \textit{The  Role  Of  Solar  And  Stellar  Energetic  Particles  On (Exo)Planetary Habitability (ETERNAL)}.

\bibliographystyle{mnras}
\bibliography{test}

\begin{thebibliography}{}
\makeatletter
\relax
\def\mn@urlcharsother{\let\do\@makeother \do\$\do\&\do\#\do\^\do\_\do\%\do\~}
\def\mn@doi{\begingroup\mn@urlcharsother \@ifnextchar [ {\mn@doi@}
  {\mn@doi@[]}}
\def\mn@doi@[#1]#2{\def\@tempa{#1}\ifx\@tempa\@empty \href
  {http://dx.doi.org/#2} {doi:#2}\else \href {http://dx.doi.org/#2} {#1}\fi
  \endgroup}
\def\mn@eprint#1#2{\mn@eprint@#1:#2::\@nil}
\def\mn@eprint@arXiv#1{\href {http://arxiv.org/abs/#1} {{\tt arXiv:#1}}}
\def\mn@eprint@dblp#1{\href {http://dblp.uni-trier.de/rec/bibtex/#1.xml}
  {dblp:#1}}
\def\mn@eprint@#1:#2:#3:#4\@nil{\def\@tempa {#1}\def\@tempb {#2}\def\@tempc
  {#3}\ifx \@tempc \@empty \let \@tempc \@tempb \let \@tempb \@tempa \fi \ifx
  \@tempb \@empty \def\@tempb {arXiv}\fi \@ifundefined
  {mn@eprint@\@tempb}{\@tempb:\@tempc}{\expandafter \expandafter \csname
  mn@eprint@\@tempb\endcsname \expandafter{\@tempc}}}

\bibitem[\protect\citeauthoryear{{Arthur}}{{Arthur}}{2012}]{Arthur-2012}
{Arthur} S.~J.,  2012, \mn@doi [\mnras] {10.1111/j.1365-2966.2011.20388.x},
  \href {http://adsabs.harvard.edu/abs/2012MNRAS.421.1283A} {421, 1283}

\bibitem[\protect\citeauthoryear{{Benaglia}, {Romero}, {Mart{\'{\i}}}, {Peri}
  \& {Araudo}}{{Benaglia} et~al.}{2010}]{Benaglia-etal-2010}
{Benaglia} P.,  {Romero} G.~E.,  {Mart{\'{\i}}} J.,  {Peri} C.~S.,   {Araudo}
  A.~T.,  2010, \mn@doi [\aap] {10.1051/0004-6361/201015232}, \href
  {http://cdsads.u-strasbg.fr/abs/2010A%26A...517L..10B} {517, L10}

\bibitem[\protect\citeauthoryear{{Biskamp}}{{Biskamp}}{2008}]{Biskamp-2008}
{Biskamp} D.,  2008, {Magnetohydrodynamic Turbulence, Cambrigde University
  Press}

\bibitem[\protect\citeauthoryear{{Brighenti} \& {D'Ercole}}{{Brighenti} \&
  {D'Ercole}}{1997}]{Brighenti-DErcole-1997}
{Brighenti} F.,  {D'Ercole} A.,  1997, \mn@doi [\mnras]
  {10.1093/mnras/285.2.387}, \href
  {https://ui.adsabs.harvard.edu/abs/1997MNRAS.285..387B} {285, 387}

\bibitem[\protect\citeauthoryear{{Burlaga}}{{Burlaga}}{1995}]{Burlaga-1995}
{Burlaga} L.~F.,  1995, Interplanetary magnetohydrodynamics, by L.~F.~Burlag.~
  International Series in Astronomy and Astrophysics, Vol.~3, Oxford University
  Press.~1995.~272 pages; ISBN13: 978-0-19-508472-6, \href
  {http://esoads.eso.org/abs/1995ISAA....3.....B} {3}

\bibitem[\protect\citeauthoryear{{Comeron} \& {Kaper}}{{Comeron} \&
  {Kaper}}{1998}]{Comeron-Kaper-1998}
{Comeron} F.,  {Kaper} L.,  1998, \aap, 338, 273

\bibitem[\protect\citeauthoryear{{Cox} et~al.,}{{Cox}
  et~al.}{2012}]{Cox-etal-2012}
{Cox} N.~L.~J.,  et~al., 2012, \mn@doi [\aap] {10.1051/0004-6361/201117910e},
  \href {http://esoads.eso.org/abs/2012A%26A...543C...1C} {543, C1}

\bibitem[\protect\citeauthoryear{{De Becker}, {del Valle}, {Romero}, {Peri}  \&
  {Benaglia}}{{De Becker} et~al.}{2017}]{DeBecker-etal-2017}
{De Becker} M.,  {del Valle} M.~V.,  {Romero} G.~E.,  {Peri} C.~S.,
  {Benaglia} P.,  2017, \mn@doi [\mnras] {10.1093/mnras/stx1826}, \href
  {http://esoads.eso.org/abs/2017MNRAS.471.4452D} {471, 4452}

\bibitem[\protect\citeauthoryear{{Decin} et~al.,}{{Decin}
  et~al.}{2012}]{Decin-etal-2012}
{Decin} L.,  et~al., 2012, \mn@doi [\aap] {10.1051/0004-6361/201219792}, \href
  {http://esoads.eso.org/abs/2012A%26A...548A.113D} {548, A113}

\bibitem[\protect\citeauthoryear{{Gnat} \& {Ferland}}{{Gnat} \&
  {Ferland}}{2012}]{Gnat-Ferland-2012}
{Gnat} O.,  {Ferland} G.~J.,  2012, \mn@doi [\apjs]
  {10.1088/0067-0049/199/1/20}, \href
  {http://esoads.eso.org/abs/2012ApJS..199...20G} {199, 20}

\bibitem[\protect\citeauthoryear{{Goedbloed} \& {Poedts}}{{Goedbloed} \&
  {Poedts}}{2004}]{Goedbloed-Poedts-2004}
{Goedbloed} J.~P.~H.,  {Poedts} S.,  2004, {Principles of
  Magnetohydrodynamics}.
Cambridge University Press

\bibitem[\protect\citeauthoryear{{Goedbloed}, {Keppens}  \&
  {Poedts}}{{Goedbloed} et~al.}{2010}]{Goedbloed-etal-2010}
{Goedbloed} J.~P.,  {Keppens} R.,   {Poedts} S.,  2010, {Advanced
  Magnetohydrodynamics}.
Cambridge, UK: Cambridge University Press

\bibitem[\protect\citeauthoryear{{Green}, {Mackey}, {Haworth}, {Gvaramadze}  \&
  {Duffy}}{{Green} et~al.}{2019}]{Green-etal-2019}
{Green} S.,  {Mackey} J.,  {Haworth} T.~J.,  {Gvaramadze} V.~V.,   {Duffy} P.,
  2019, \mn@doi [\aap] {10.1051/0004-6361/201834832}, \href
  {https://ui.adsabs.harvard.edu/abs/2019A&A...625A...4G} {625, A4}

\bibitem[\protect\citeauthoryear{{Gvaramadze} \& {Bomans}}{{Gvaramadze} \&
  {Bomans}}{2008}]{Gvaramadze-Bomans-2008}
{Gvaramadze} V.~V.,  {Bomans} D.~J.,  2008, \mn@doi [\aap]
  {10.1051/0004-6361:200810411}, 490, 1071

\bibitem[\protect\citeauthoryear{{Gvaramadze}, {Kniazev}, {Kroupa}  \&
  {Oh}}{{Gvaramadze} et~al.}{2011}]{Gvaramadze-etal-2011}
{Gvaramadze} V.~V.,  {Kniazev} A.~Y.,  {Kroupa} P.,   {Oh} S.,  2011, \mn@doi
  [\aap] {10.1051/0004-6361/201117746}, \href
  {http://esoads.eso.org/abs/2011A%26A...535A..29G} {535, A29}

\bibitem[\protect\citeauthoryear{{Gvaramadze}, {Alexashov}, {Katushkina}  \&
  {Kniazev}}{{Gvaramadze} et~al.}{2018}]{Gvaramadze-etal-2018}
{Gvaramadze} V.~V.,  {Alexashov} D.~B.,  {Katushkina} O.~A.,   {Kniazev} A.~Y.,
   2018, \mn@doi [\mnras] {10.1093/mnras/stx3089}, \href
  {https://ui.adsabs.harvard.edu/abs/2018MNRAS.474.4421G} {474, 4421}

\bibitem[\protect\citeauthoryear{{Ha}, {Ryu}, {Kang}  \& {van Marle}}{{Ha}
  et~al.}{2018}]{Ha-etal-2018}
{Ha} J.-H.,  {Ryu} D.,  {Kang} H.,   {van Marle} A.~J.,  2018, \mn@doi [\apj]
  {10.3847/1538-4357/aad634}, \href
  {http://esoads.eso.org/abs/2018ApJ...864..105H} {864, 105}

\bibitem[\protect\citeauthoryear{{Huthoff} \& {Kaper}}{{Huthoff} \&
  {Kaper}}{2002}]{Huthoff-Kaper-2002}
{Huthoff} F.,  {Kaper} L.,  2002, \mn@doi [\aap] {10.1051/0004-6361:20011793},
  \href {http://esoads.eso.org/abs/2002A\%26A...383..999H} {383, 999}

\bibitem[\protect\citeauthoryear{{Katushkina}, {Alexashov}, {Izmodenov}  \&
  {Gvaramadze}}{{Katushkina} et~al.}{2017}]{Katushkina-etal-2017}
{Katushkina} O.~A.,  {Alexashov} D.~B.,  {Izmodenov} V.~V.,   {Gvaramadze}
  V.~V.,  2017, \mn@doi [\mnras] {10.1093/mnras/stw2833}, \href
  {https://ui.adsabs.harvard.edu/abs/2017MNRAS.465.1573K} {465, 1573}

\bibitem[\protect\citeauthoryear{{Katushkina}, {Alexashov}, {Gvaramadze}  \&
  {Izmodenov}}{{Katushkina} et~al.}{2018}]{Katushkina-etal-2018}
{Katushkina} O.~A.,  {Alexashov} D.~B.,  {Gvaramadze} V.~V.,   {Izmodenov}
  V.~V.,  2018, \mn@doi [\mnras] {10.1093/mnras/stx2488}, \href
  {https://ui.adsabs.harvard.edu/abs/2018MNRAS.473.1576K} {473, 1576}

\bibitem[\protect\citeauthoryear{{Kissmann}, {Kleimann}, {Krebl}  \&
  {Wiengarten}}{{Kissmann} et~al.}{2018}]{Kissmann-etal-2018}
{Kissmann} R.,  {Kleimann} J.,  {Krebl} B.,   {Wiengarten} T.,  2018, \mn@doi
  [\apjs] {10.3847/1538-4365/aabe75}, \href
  {http://adsabs.harvard.edu/abs/2018ApJS..236...53K} {236, 53}

\bibitem[\protect\citeauthoryear{{Kobulnicky}, {Gilbert}  \&
  {Kiminki}}{{Kobulnicky} et~al.}{2010}]{Kobulnicky-etal-2010}
{Kobulnicky} H.~A.,  {Gilbert} I.~J.,   {Kiminki} D.~C.,  2010, \mn@doi [\apj]
  {10.1088/0004-637X/710/1/549}, 710, 549

\bibitem[\protect\citeauthoryear{{Kobulnicky}, {Schurhammer}, {Baldwin},
  {Chick}, {Dixon}, {Lee}  \& {Povich}}{{Kobulnicky}
  et~al.}{2017}]{Kobulnicky-etal-2017}
{Kobulnicky} H.~A.,  {Schurhammer} D.~P.,  {Baldwin} D.~J.,  {Chick} W.~T.,
  {Dixon} D.~M.,  {Lee} D.,   {Povich} M.~S.,  2017, \mn@doi [\aj]
  {10.3847/1538-3881/aa90ba}, \href
  {http://cdsads.u-strasbg.fr/abs/2017AJ....154..201K} {154, 201}

\bibitem[\protect\citeauthoryear{{Kosi{\'n}ski} \& {Hanasz}}{{Kosi{\'n}ski} \&
  {Hanasz}}{2006}]{Kosinski-Hanasz-2006}
{Kosi{\'n}ski} R.,  {Hanasz} M.,  2006, \mn@doi [\mnras]
  {10.1111/j.1365-2966.2006.10142.x}, \href
  {http://esoads.eso.org/abs/2006MNRAS.368..759K} {368, 759}

\bibitem[\protect\citeauthoryear{{Lamers} \& {Cassinelli}}{{Lamers} \&
  {Cassinelli}}{1999}]{Lamers-Cassinelli-1999}
{Lamers} H.~J.~G.~L.~M.,  {Cassinelli} J.~P.,  1999, {Introduction to Stellar
  Winds}.
Cambridge, UK: Cambridge University Press, ISBN 0521593980

\bibitem[\protect\citeauthoryear{{Landau} \& {Lifshitz}}{{Landau} \&
  {Lifshitz}}{1984}]{Landau-Lifshitz-1984}
{Landau} L.~D.,  {Lifshitz} E.~M.,  1984, {Electrodynamics of Continuous
  Media}.
Pergamon Press

\bibitem[\protect\citeauthoryear{{Landau} \& {Lifshitz}}{{Landau} \&
  {Lifshitz}}{1987}]{Landau-Lifshitz-1987}
{Landau} L.~D.,  {Lifshitz} E.~M.,  1987, {Fluid Mechanics}.
Pergamon Press

\bibitem[\protect\citeauthoryear{{Linsky} \& {Wood}}{{Linsky} \&
  {Wood}}{2014}]{Linsky-Wood-2014}
{Linsky} J.~L.,  {Wood} B.~E.,  2014, \mn@doi [ASTRA Proceedings]
  {10.5194/ap-1-43-2014}, \href {http://esoads.eso.org/abs/2014ASTRP...1...43L}
  {1, 43}

\bibitem[\protect\citeauthoryear{{Mackey}, {Langer}, {Mohamed}, {Gvaramadze},
  {Neilson}  \& {Meyer}}{{Mackey} et~al.}{2014}]{Mackey-etal-2014b}
{Mackey} J.,  {Langer} N.,  {Mohamed} S.,  {Gvaramadze} V.~V.,  {Neilson}
  H.~R.,   {Meyer} D.~M.-A.,  2014, \mn@doi [ASTRA Proceedings]
  {10.5194/ap-1-61-2014}, \href {http://esoads.eso.org/abs/2014ASTRP...1...61M}
  {1, 61}

\bibitem[\protect\citeauthoryear{{Mackey}, {Gvaramadze}, {Mohamed}  \&
  {Langer}}{{Mackey} et~al.}{2015}]{Mackey-etal-2015}
{Mackey} J.,  {Gvaramadze} V.~V.,  {Mohamed} S.,   {Langer} N.,  2015, \mn@doi
  [\aap] {10.1051/0004-6361/201424716}, \href
  {http://esoads.eso.org/abs/2015A%26A...573A..10M} {573, A10}

\bibitem[\protect\citeauthoryear{{Mellema} \& {Lundqvist}}{{Mellema} \&
  {Lundqvist}}{2002}]{Mellema-Lundqvist-2002}
{Mellema} G.,  {Lundqvist} P.,  2002, \mn@doi [\aap]
  {10.1051/0004-6361:20021164}, 394, 901

\bibitem[\protect\citeauthoryear{{Meyer}, {van Marle}, {Kuiper}  \&
  {Kley}}{{Meyer} et~al.}{2016}]{Meyer-etal-2016}
{Meyer} D.~M.-A.,  {van Marle} A.-J.,  {Kuiper} R.,   {Kley} W.,  2016, \mn@doi
  [\mnras] {10.1093/mnras/stw651}, \href
  {http://esoads.eso.org/abs/2016MNRAS.459.1146M} {459, 1146}

\bibitem[\protect\citeauthoryear{{Meyer}, {Mignone}, {Kuiper}, {Raga}  \&
  {Kley}}{{Meyer} et~al.}{2017}]{Meyer-etal-2017}
{Meyer} D.~M.-A.,  {Mignone} A.,  {Kuiper} R.,  {Raga} A.~C.,   {Kley} W.,
  2017, \mn@doi [\mnras] {10.1093/mnras/stw2537}, \href
  {http://esoads.eso.org/abs/2017MNRAS.464.3229M} {464, 3229}

\bibitem[\protect\citeauthoryear{{Moloto}, {Engelbrecht}  \& {Burger}}{{Moloto}
  et~al.}{2018}]{Moloto-etal-2018}
{Moloto} K.~D.,  {Engelbrecht} N.~E.,   {Burger} R.~A.,  2018, \mn@doi [\apj]
  {10.3847/1538-4357/aac174}, \href
  {https://ui.adsabs.harvard.edu/#abs/2018ApJ...859..107M} {859, 107}

\bibitem[\protect\citeauthoryear{{Opher}, {Drake}, {Velli}, {Decker}  \&
  {Toth}}{{Opher} et~al.}{2012}]{Opher-etal-2012}
{Opher} M.,  {Drake} J.~F.,  {Velli} M.,  {Decker} R.~B.,   {Toth} G.,  2012,
  \mn@doi [\apj] {10.1088/0004-637X/751/2/80}, \href
  {http://esoads.eso.org/abs/2012ApJ...751...80O} {751, 80}

\bibitem[\protect\citeauthoryear{{Parker}}{{Parker}}{1958}]{Parker-1958}
{Parker} E.~N.,  1958, \apj, 128, 664

\bibitem[\protect\citeauthoryear{{Peri}, {Benaglia}  \& {Isequilla}}{{Peri}
  et~al.}{2015}]{Peri-etal-2015}
{Peri} C.~S.,  {Benaglia} P.,   {Isequilla} N.~L.,  2015, \mn@doi [\aap]
  {10.1051/0004-6361/201424676}, \href
  {http://esoads.eso.org/abs/2015A\%26A...578A..45P} {578, A45}

\bibitem[\protect\citeauthoryear{{Planck Collaboration} et~al.,}{{Planck
  Collaboration} et~al.}{2016}]{Planck-Collaboration-2016}
{Planck Collaboration} et~al., 2016, \mn@doi [\aap]
  {10.1051/0004-6361/201528033}, \href
  {http://esoads.eso.org/abs/2016A%26A...596A.103P} {596, A103}

\bibitem[\protect\citeauthoryear{{Pogorelov} et~al.,}{{Pogorelov}
  et~al.}{2017a}]{Pogorelov-etal-2017}
{Pogorelov} N.~V.,  et~al., 2017a, \mn@doi [\ssr] {10.1007/s11214-017-0354-8},
  \href {http://esoads.eso.org/abs/2017SSRv..tmp...29P} {}

\bibitem[\protect\citeauthoryear{{Pogorelov}, {Heerikhuisen}, {Roytershteyn},
  {Burlaga}, {Gurnett}  \& {Kurth}}{{Pogorelov}
  et~al.}{2017b}]{Pogorelov-etal-2017b}
{Pogorelov} N.~V.,  {Heerikhuisen} J.,  {Roytershteyn} V.,  {Burlaga} L.~F.,
  {Gurnett} D.~A.,   {Kurth} W.~S.,  2017b, \mn@doi [\apj]
  {10.3847/1538-4357/aa7d4f}, \href
  {http://esoads.eso.org/abs/2017ApJ...845....9P} {845, 9}

\bibitem[\protect\citeauthoryear{{Reyes-Iturbide}, {Vel{\'a}zquez}, {Rosado},
  {Mat{\'\i}as Schneiter}  \& {Ram{\'\i}rez-Ballinas}}{{Reyes-Iturbide}
  et~al.}{2019}]{Reyes-Iturbide-etal-2019}
{Reyes-Iturbide} J.,  {Vel{\'a}zquez} P.~F.,  {Rosado} M.,  {Mat{\'\i}as
  Schneiter} E.,   {Ram{\'\i}rez-Ballinas} I.,  2019, \mn@doi [\rmxaa]
  {10.22201/ia.01851101p.2019.55.02.09}, \href
  {https://ui.adsabs.harvard.edu/abs/2019RMxAA..55..211R} {55, 211}

\bibitem[\protect\citeauthoryear{{Ruderman} \& {Brevdo}}{{Ruderman} \&
  {Brevdo}}{2006}]{Ruderman-Brevdo-2006}
{Ruderman} M.~S.,  {Brevdo} L.,  2006, \mn@doi [\aap]
  {10.1051/0004-6361:20053854}, \href
  {https://ui.adsabs.harvard.edu/abs/2006A&A...448.1177R} {448, 1177}

\bibitem[\protect\citeauthoryear{{Scherer}, {Fichtner}, {Fahr}  \&
  {R\"oken}}{{Scherer} et~al.}{2015a}]{Scherer-etal-2015c}
{Scherer} K.,  {Fichtner} H.,  {Fahr} H.-J.,   {R\"oken} C.,  2015a, Submitted
  to \apj

\bibitem[\protect\citeauthoryear{{Scherer} et~al.,}{{Scherer}
  et~al.}{2015b}]{Scherer-etal-2015a}
{Scherer} K.,  et~al., 2015b, \mn@doi [\aap] {10.1051/0004-6361/201425091},
  \href {http://esoads.eso.org/abs/2015A\%26A...576A..97S} {576, A97}

\bibitem[\protect\citeauthoryear{{Scherer}, {Strauss}, {Ferreira}  \&
  {Fichtner}}{{Scherer} et~al.}{2016a}]{Scherer-etal-2016c}
{Scherer} K.,  {Strauss} R.~D.,  {Ferreira} S.~E.~S.,   {Fichtner} H.,  2016a,
  \mn@doi [Astroparticle Physics] {10.1016/j.astropartphys.2016.06.003}, \href
  {http://esoads.eso.org/abs/2016APh....82...93S} {82, 93}

\bibitem[\protect\citeauthoryear{{Scherer}, {Fichtner}, {Kleimann},
  {Wiengarten}, {Bomans}  \& {Weis}}{{Scherer}
  et~al.}{2016b}]{Scherer-etal-2016a}
{Scherer} K.,  {Fichtner} H.,  {Kleimann} J.,  {Wiengarten} T.,  {Bomans}
  D.~J.,   {Weis} K.,  2016b, \mn@doi [\aap] {10.1051/0004-6361/201526137},
  \href {http://esoads.eso.org/abs/2016A%26A...586A.111S} {586, A111}

\bibitem[\protect\citeauthoryear{{Schlickeiser}}{{Schlickeiser}}{2002}]{Schlickeiser-2002}
{Schlickeiser} R.,  2002, {Cosmic Ray Astrophysics}.
Astronomy and Astrophysics Library, Springer, Berlin.~ISBN 3-540-66465-3

\bibitem[\protect\citeauthoryear{{Schure}, {Kosenko}, {Kaastra}, {Keppens}  \&
  {Vink}}{{Schure} et~al.}{2009}]{Schure-etal-2009}
{Schure} K.~M.,  {Kosenko} D.,  {Kaastra} J.~S.,  {Keppens} R.,   {Vink} J.,
  2009, \mn@doi [\aap] {10.1051/0004-6361/200912495}, \href
  {http://esoads.eso.org/abs/2009A%26A...508..751S} {508, 751}

\bibitem[\protect\citeauthoryear{{Sutherland} \& {Dopita}}{{Sutherland} \&
  {Dopita}}{1993}]{Sutherland-Dopita-1993}
{Sutherland} R.~S.,  {Dopita} M.~A.,  1993, \mn@doi [\apjs] {10.1086/191823},
  \href {http://esoads.eso.org/abs/1993ApJS...88..253S} {88, 253}

\bibitem[\protect\citeauthoryear{{Walder}, {Folini}  \& {Meynet}}{{Walder}
  et~al.}{2012}]{Walder-etal-2012}
{Walder} R.,  {Folini} D.,   {Meynet} G.,  2012, \mn@doi [\ssr]
  {10.1007/s11214-011-9771-2}, \href
  {http://esoads.eso.org/abs/2012SSRv..166..145W} {166, 145}

\bibitem[\protect\citeauthoryear{{Wang} \& {Belcher}}{{Wang} \&
  {Belcher}}{1998}]{Wang-Belcher-1998}
{Wang} C.,  {Belcher} J.~W.,  1998, \mn@doi [\jgr] {10.1029/97JA02773}, \href
  {https://ui.adsabs.harvard.edu/abs/1998JGR...103..247W} {103, 247}

\bibitem[\protect\citeauthoryear{{Webb}, {Drury}  \& {Volk}}{{Webb}
  et~al.}{1986}]{Webb-etal-1986}
{Webb} G.~M.,  {Drury} L.~O.,   {Volk} H.~J.,  1986, \aap, \href
  {https://ui.adsabs.harvard.edu/#abs/1986A&A...160..335W} {160, 335}

\bibitem[\protect\citeauthoryear{{Wilkin}}{{Wilkin}}{2000}]{Wilkin-2000}
{Wilkin} F.~P.,  2000, \mn@doi [\apj] {10.1086/308576}, \href
  {http://esoads.eso.org/abs/2000ApJ...532..400W} {532, 400}

\bibitem[\protect\citeauthoryear{{Wood}, {Izmodenov}, {Linsky}  \&
  {Alexashov}}{{Wood} et~al.}{2007}]{Wood-etal-2007}
{Wood} B.~E.,  {Izmodenov} V.~V.,  {Linsky} J.~L.,   {Alexashov} D.,  2007,
  \mn@doi [\apj] {10.1086/512482}, 659, 1784

\bibitem[\protect\citeauthoryear{{Zank}, {Heerikhuisen}, {Wood}, {Pogorelov},
  {Zirnstein}  \& {McComas}}{{Zank} et~al.}{2013}]{Zank-etal-2013}
{Zank} G.~P.,  {Heerikhuisen} J.,  {Wood} B.~E.,  {Pogorelov} N.~V.,
  {Zirnstein} E.,   {McComas} D.~J.,  2013, \mn@doi [\apj]
  {10.1088/0004-637X/763/1/20}, \href
  {http://esoads.eso.org/abs/2013ApJ...763...20Z} {763, 20}

\bibitem[\protect\citeauthoryear{{del Valle} \& {Pohl}}{{del Valle} \&
  {Pohl}}{2018}]{delValle-Pohl-2018}
{del Valle} M.~V.,  {Pohl} M.,  2018, \mn@doi [\apj]
  {10.3847/1538-4357/aad333}, \href
  {http://cdsads.u-strasbg.fr/abs/2018ApJ...864...19D} {864, 19}

\bibitem[\protect\citeauthoryear{{van Marle}, {Decin}, {Cox}  \&
  {Meliani}}{{van Marle} et~al.}{2014a}]{vanMarle-etal-2014}
{van Marle} A.~J.,  {Decin} L.,  {Cox} N.,   {Meliani} Z.,  2014a, preprint,
  \href {http://esoads.eso.org/abs/2014arXiv1407.1620V} {} (\mn@eprint {arXiv}
  {1407.1620})

\bibitem[\protect\citeauthoryear{{van Marle}, {Decin}  \& {Meliani}}{{van
  Marle} et~al.}{2014b}]{van-Marle-etal-2014}
{van Marle} A.~J.,  {Decin} L.,   {Meliani} Z.,  2014b, \mn@doi [\aap]
  {10.1051/0004-6361/201321968}, \href
  {http://esoads.eso.org/abs/2014A%26A...561A.152V} {561, A152}

\bibitem[\protect\citeauthoryear{{van Marle}, {Decin}, {Cox}  \&
  {Meliani}}{{van Marle} et~al.}{2015}]{vanMarle-etal-2015}
{van Marle} A.~J.,  {Decin} L.,  {Cox} N.~L.~J.,   {Meliani} Z.,  2015, \mn@doi
  [Journal of Physics Conference Series] {10.1088/1742-6596/577/1/012024},
  \href {http://esoads.eso.org/abs/2015JPhCS.577a2024V} {577, 012024}

\makeatother
\end{thebibliography}

\appendix
\section{The Rankine-Hugoniot relations}
To obtain the Rankine-Hugoniot relation in (magneto-)hydro\-dy\-namics, one shall use the conservative form of the Euler equations, in which one then can replace the divergence by the multiplication with the normal vector and the partial time derivative by multiplication with the negative shock speed ($-u$) \citep{Goedbloed-Poedts-2004,Goedbloed-etal-2010}. In the general case (oblique shocks), it is hard to determine the shock normal, while from models it is relatively easy to do: Knowing the upstream and downstream magnetic fields (or velocities for the HD case) one can use the coplanarity theorem \citep{Burlaga-1995}:
\begin{align}
    \vec{n} &= \frac{(\vec{B}_{1}-\vec{B}_{2}) \times (\vec{B}_{1}\times\vec{B}_{2})}
    {\left|(\vec{B}_{1}-\vec{B}_{2}) \times (\vec{B}_{1}\times\vec{B}_{2})\right|}
\end{align}
or in the HD case
\begin{align}
    \vec{n} &= \frac{\vec{v}_{2}-\vec{v}_{1}}
              {|\vec{v}_{2}-\vec{v}_{1}|}   
\end{align}
We denote with the index $\mathrm{n}$ the projection of a vector to the normal, for example $v_{\mathrm{n}} = \vec{v}\cdot\vec{n}$, hence $\vec{v}_{\mathrm{t}}=\vec{v}-v_{\mathrm{n}}\vec{n}$.  Note that there are two tangential directions
perpendicular to each other.  

The indices $1,2$ denote the upstream and downstream region, respectively. To simplify the calculations, one either transforms into the shock rest frame (i.e.\ $\vec{v}' = \vec{v}- u\vec{n}$), or assumes that the shock is stationary $u=0$ (we neglect the $'$ in the following). The indices $\mathrm{n,t}$ then denote the normal and tangential
components of the vectors, respectively; $\rho,P, \vec{u},\vec{B}, \gamma$ are the
density, thermal pressure, plasma (bulk) velocity, magnetic field and
polytropic index. We introduce the following short hand notations:

%\scalebox{0.85}{\begin{tabular}{lrll}
\begin{align*}
  v_{\mathrm{A,n},i}&\equiv\dfrac{B_{\mathrm{n},i}}{\sqrt{\rho}} &  \text{normal Alfv\'en  speed}\\
  v_{\mathrm{c}}    &\equiv\sqrt{\dfrac{\gamma P}{\rho}} & \text{sound speed}\\
  M&\equiv M_{\mathrm{A,n},i}\equiv  \dfrac{u_{\mathrm{n},i}}{V_{\mathrm{A,n},i}} & \text{normal Alfv\'enic Mach number}\\
  M_{\mathrm{A,t},i}&\equiv\dfrac{u_{\mathrm{n},1}\sqrt{\rho_{i}}}{B_{\mathrm{t},i}} & \text{``tangential'' Alfv\'enic Mach  number}\\
  M_{\mathrm{s,n},i}&\equiv\dfrac{u_{\mathrm{n}}}{v_{\mathrm{c}}} & \text{normal sonic Mach number}\\
  s&\equiv\dfrac{\rho_{2}}{\rho_{1}}=\dfrac{u_{\mathrm{n},1}}{u_{\mathrm{n},2}}& \text{compression ratio}
  \end{align*}
%\end{tabular}}
~\\
The normalized MHD Rankine-Hugoniot equations are \citep[e.g.][]{Goedbloed-etal-2010}

%\onecolumn
\begin{subequations}
    \label{eq:ms3}
  \begin{align}
     \label{eq:ms3a}
      \rhb{\rho u_{\mathrm{n}}} &=0 %&\text{(continuity)}
      &\\
    \label{eq:ms3b}
\rhb{\rho u_{\mathrm{n}}^{2} +P + \dfrac{1}{2} B^{2}_{\mathrm{t}}} &=0 %&\text{(normal momentum)} 
&\\
    \label{eq:ms3c}
\rho u_{\mathrm{n}}\rhb{\vec{u}_{\mathrm{t}}} - B_{\mathrm{n}}\rhb{\vec{B}_{\mathrm{t}}} &=0 %&\text{(tangential momentum)}
&\\ \nonumber
\rho u_{\mathrm{n}}\rhb{\dfrac{1}{2} (u^{2}_{\mathrm{t}}+u_{\mathrm{n}}^{2}) +
    \dfrac{1}{\rho}\left(\dfrac{\gamma}{\gamma-1} P+B^{2}_{\mathrm{t}}\right)}
  - &\\ \label{eq:ms3d}
  -B_{\mathrm{n}}\rhb{(\vec{u}_{\mathrm{t}}\cdot\vec{B}_{\mathrm{t}})} &=0 %&\text{(energy)}
  &\\
    \label{eq:ms3e}
\rhb{B_{\mathrm{n}}} &=0 %&\text{(normal $B$-flux)} 
&\\
    \label{eq:ms3f}
\rho u_{\mathrm{n}} \rhb{\dfrac{\vec{B}_{\mathrm{t}}}{\rho}}-B_{\mathrm{n}}\rhb{\vec{u}_{\mathrm{t}}}&= 0
                           %&\text{(tangential $B$-flux)} 
  \end{align}
\end{subequations}

If $\vec{B}=\vec{0}$, we have a hydrodynamic shock, for $B_{\mathrm{n}}$ a perpendicular shock, and for  $B_{\mathrm{n}}\ne0, \vec{B}_{\mathrm{t}}\ne\vec{0}$ a genuine (oblique) shock.
Then the  solution of the Eqs.~(\ref{eq:ms3}) are in the HD case linear, for perpendicular shocks quadratic, and for genuine shocks cubic in $s$ . The latter can be found after a tedious but straightforward algebraic manipulation. Thus we have 
\begin{align}
    a_{3}s^{3}+a_{2}s^{2}+a_{1}s^{1}+a_{0}= 0
\end{align}

The coefficients for the HD case are:
\begin{subequations}
\label{HD}
    \begin{align}
        a_{3} &= 0\\\nonumber
        a_{2} &= 0\\\nonumber
        a_{1} &= 2 + (\gamma - 1) M_{\mathrm{s,n},1}^2\\\nonumber
        a_{0} &= -(\gamma+1) \Msu\\\nonumber
    \end{align}
\end{subequations}
for the perpendicular shocks:
\begin{subequations}
\label{PS}
    \begin{align}
        a_{3} &= 0\\\nonumber
        a_{2} &= 2\Msu\\\nonumber
        a_{1} &= (2+(\gamma-1)\Msu)M_{\mathrm{A.t},1}^{2} + 2\gamma \Msu\\\nonumber
        a_{0} &= -(\gamma+1) M_{\mathrm{A.t},1}^{2}\Msu\\\nonumber
    \end{align}
\end{subequations}
and for the genuine shocks:
%\begin{strip}
\begin{subequations}
\label{GS}
    \begin{align}
        a_{3} &= \left[2-(\gamma-1)\left(2 D - 1
          +3\tant\right)\Msu\right]\M\\\nonumber
          &\hspace*{1cm}-2 (\gamma-1)(D-2\tant) \Msu
\\\nonumber
        a_{2} &= \left[-4+\left((2D -2-\tant)(\gamma-1)+\tant\right)\Msu 
          \right]\Mq - \\\nonumber    
  &\hspace*{1cm}\left[(\tant+1)(\gamma+1) - 2D(\gamma-1)\right]\Msu\M\\\nonumber
        a_{1} &= \left[2 + (\gamma -1) \Msu\right]\Msq +
          \left[\gamma(\tant+2)+2\right]\Msu\Mq\\\nonumber
        a_{0} &=  -(\gamma+1)\Msu\Msq
    \end{align}
with
\begin{align}
    D=
         \dfrac{\vec{u}_{\mathrm{t},1}\cdot\vec{B}_{\mathrm{t},1}}{u_{\mathrm{n},1}B_{\mathrm{n}}} \ .
\end{align}
\end{subequations}
%\end{strip}

Having determined the compression ratio $s$, we can easily calculate the remaining parameters of interest from Eqs.~(\ref{eq:ms3}). For the perpendicular and HD shocks we have Table~\ref{tab:a1} and for the genuine shocks Table~\ref{tab:a2}.

\begin{table}
  \caption{\label{tab:1}Compression ratios for
    HD and perpendicular shocks. In the HD case the last three lines are not defined and $M_{\mathrm{A,t},1}\to\infty$.}\label{tab:a1}
  \begin{tabular}{RLX}\toprule
     \multicolumn{2}{l}{exact}&\multicolumn{1}{l}{$\lim\limits_{M_{\mathrm{A,t},1}\to\infty}$ }\\
\midrule
% rho
    \dfrac{\rho_{2}}{\rho_{1}}
   &\dfrac{u_{\mathrm{n},1}}{u_{\mathrm{n},2}}  =s
   &     s\\ \addlinespace
   % tangential velocity
     \vec{u}_{\mathrm{t},2} & \vec{u}_{\mathrm{t},1}&-\\ \addlinespace
     %pressure
\dfrac{P_{2}}{P_{1}} &  \gamma \left(1-\dfrac{1}{s}\right)\Msu 
         +\dfrac{\gamma}{2}\frac{\Msu}{M_{\mathrm{A,t},1}^2}(1-s^{2}) + 1&\gamma \left(1-\dfrac{1}{s}\right)\Msu \\ \addlinespace
         %modulus of tangential field
           \dfrac{B_{\mathrm{t},2}}{B_{\mathrm{t},2}}  & s&-\\ \addlinespace
           %\tangential field vector
             \vec{B}_{\mathrm{t},2}&s\vec{B}_{\mathrm{t},1}&-\\ \addlinespace
  \dfrac{v_{\mathrm{A,t},2}^{2}}{v_{\mathrm{A,t},1}^{2}}  & s& -\\ \bottomrule
  \end{tabular}
\end{table}

\begin{table}
  \caption{\label{tab:v1}Compression ratios and downstream values for
    vectors for the genuine shocks. }\label{tab:a2}
  \begin{tabular}{RLX}
    \toprule
     \multicolumn{2}{l}{exact}&\multicolumn{1}{l}{$\lim\limits_{\M\to\infty}$ }\\
\midrule
% rho
    \dfrac{\rho_{2}}{\rho_{1}}
   &\dfrac{u_{\mathrm{n},1}}{u_{\mathrm{n},2}}  =s
   &     s\\ \addlinespace
%u &
    \vec{u}_{\mathrm{t},2}
   &  \vec{u}_{\mathrm{t},1} +\dfrac{(1-s)}{s-\M}\,\dfrac{u_{\mathrm{n},1}}{B_{\mathrm{n}}}\vec{B}_{\mathrm{t},1}\
   & \vec{u}_{\mathrm{t},1}\\ \addlinespace
% &
    \dfrac{P_{2}}{P_{1}}
   & \gamma \Msu\dfrac{s-1}{s}+ 1 +
     \dfrac{(1-s)\M}{\beta_{\mathrm{t},1}}\,\dfrac{(s+1)\M-2s}{(\M-s)^{2}}
   &   \gamma \Msu\dfrac{s-1}{s}+ 1 - \frac{1-s^{2}}{\beta_{\mathrm{t},1}} \\ \addlinespace
%B &
     \vec{B}_{\mathrm{t},2}
   & s \left[ \dfrac{\M-1}{\M-s}\right]\vec{B}_{\mathrm{t},1}
   &s\vec{B}_{\mathrm{t},1}\\ \addlinespace
%  & B2/B1    
     B_{2}^{2}
   & s^{2}
     \dfrac{(\M-1)^{2}}{(\M-s)^{2}}B_{1}^{2}-(s-1)\M\dfrac{(s+1)\M-2s}{(s-\M)^{2}}B_{\mathrm{n}}^{2}
   & s^{2}B_{\mathrm{t},1}^{2}-(s^{2}-1)B_{\mathrm{n}}^{2}\\ \addlinespace
%  & 
     \dfrac{B_{2}^{2}}{B_{1}^{2}}
   &  s^{2}  \dfrac{(\M-1)^{2}}{(\M-s)^{2}}
     -(s-1)\M\dfrac{(s+1)\M-2s}{(s-\M)^{2}}\cos^{2}\vartheta_{1}
   &s^{2} \sin^{2}\vartheta_{1}+\cos^{2}\vartheta_{1}\\ \addlinespace
% &
    \dfrac{v_{\mathrm{A,n},2}^{2}}{v_{\mathrm{A,n},1}^{2}}
   &\dfrac{1}{s} 
   &\dfrac{1}{s} \\ \addlinespace
% &
    \dfrac{v_{\mathrm{A,t},2}^{2}}{v_{\mathrm{A,t},1}^{2}}
   & s\dfrac{(\M-1)^{2}}{(\M-s)^{2}}
   & s\\ \bottomrule
  \end{tabular}
\end{table}

With the compression ratio and the ideal gas law we get easily the temperature, and all other parameters in mind.
\label{lastpage}
\end{document}